\documentclass[fleqn,usenatbib,useAMS]{mnras}

\usepackage{amssymb,amsmath,graphicx,natbib,hyperref}
\usepackage{color}
\usepackage[T1]{fontenc}
\usepackage{ae,aecompl}
\usepackage{graphicx}	
\usepackage{amsmath}	
\usepackage{amssymb}	
\usepackage{multicol}       
\usepackage{bm}		
\usepackage{pdflscape}	
\usepackage{newtxtext,newtxmath}

\begin{document}
\newcommand{\bibtex}{\textsc{Bib}\!\TeX} 
\newcommand{\hi}          {\mbox{\rm H{\small I}}}
\newcommand{\HI}          {\hi}
\newcommand{\hii}         {\mbox{\rm H{\small II}}}
\newcommand{\htwo}        {\mbox{H$_{2}$}}
\newcommand{\Sgas}        {\mbox{$\Sigma_{\rm gas}$}\,}
\newcommand{\Sstar}       {\mbox{$\Sigma_*$}}
\newcommand{\Ssfr}        {\mbox{$\Sigma_{\rm SFR}$}\,}
\newcommand{\Shtwo}         {\mbox{$\Sigma_{\rm H_2}$}\,}
\newcommand{\av}          {\mbox{${\rm A_V}$\,}}
\newcommand{\Ico}         {\mbox{I$_{\rm CO}$}\,}
\newcommand{\aco}         {\mbox{$\alpha_{\rm CO}$}}
\newcommand{\ha}          {\mbox{H$\alpha$}\,}
\newcommand{\hb}          {\mbox{H$\beta$}\,}
\newcommand{\msunperpcsq} {\mbox{\rm M$_\odot$~pc$^{-2}$}}
\newcommand{\sigavUnits}  {\mbox{$\mathrm{M_{\odot} \,pc^{-2}
\,mag^{-1}}$}}
\newcommand{\acounits}  {\mbox{\rm M$_\odot$ (K km s$^{-1}$ pc$^2$)$^{-1}$}}
\newcommand{\Kkmpers}     {\mbox{\rm K km s$^{-1}$}}
\newcommand{\sighi}{\mbox{$\Sigma_{\rm HI}$}\,}
\newcommand{\Reff}       {\mbox {${\rm R_{eff}}$}\,}
\newcommand{\msunperpcsqpermag} {\mbox{\rm M$_\odot$~pc$^{-2}$~mag$^{-1}$}}
\newcommand{\msun}        {\mbox{\rm M$_\odot$}}
\newcommand{\xcounits}    {\mbox{\rm cm$^{-2}$(K km s$^{-1}$)$^{-1}$}}
\newcommand{\xco}         {\mbox{$X_{\rm CO}$}}
\newcommand{\nii}         {\mbox{\rm [N{\small II}]}}

\title[]{The EDGE-CALIFA Survey: Using Optical Extinction to Probe the Spatially-Resolved Distribution of Gas in Nearby Galaxies}

\author[Barrera-Ballesteros et al.]{Jorge K. Barrera-Ballesteros$^{1}$\thanks{E-mail:jkbarrerab@astro.unam.mx},
          Dyas Utomo$^{2}$,
          Alberto D. Bolatto$^{3}$,
          \newauthor
          Sebasti\'{a}n F. S\'{a}nchez$^{1}$,
          Stuart N. Vogel$^{3}$,
          Tony Wong $^{4}$,
          Rebecca C. Levy$^{3}$,
          Dario Colombo$^{5}$,
          \newauthor
          Veselina Kalinova$^{5}$,
          Peter Teuben$^{3}$,
          Rub\'{e}n Garc\'{i}a-Benito$^{6}$,
          Bernd Husemann$^{7}$,
          \newauthor
          Dami\'{a}n Mast$^{8,9}$, and
          Leo Blitz$^{10}$
\\
$^{1}$ Instituto de Astronom\'{i}a, Universidad Nacional Aut\'{o}noma de M\'{e}xico, A.P. 70-264, 04510 M\'{e}xico, D.F.,  Mexico \\
$^{2}$ Department of Astronomy, The Ohio State University, Columbus, OH 43210, USA \\
$^{3}$ Department of Astronomy, University of Maryland, College Park, MD 20742, USA \\
$^{4}$ Department of Astronomy, University of Illinois, Urbana, IL 61801, USA \\
$^{5}$ Max-Planck-Institut f\"{u}r Radioastronomie, D-53121, Bonn, Germany \\
$^{6}$ Instituto de Astrof\'{i}sica de Andaluc\'{i}a, CSIC, E-18008 Granada, Spain \\
$^{7}$ Max-Planck-Institut f\"{u}r Astronomie, K\"onigstuhl 17, D-69117 Heidelberg, Germany \\
$^{8}$ Universidad Nacional de C\'{o}rdoba, Observatorio Astron\'{o}mico de C\'{o}rdoba, C\'{o}rdoba, Argentina \\
$^{9}$ Consejo de Investigaciones Cient\'{i}ficas y T\'{e}cnicas de la Rep\'{u}blica Argentina, C1033AAJ, CABA, Argentina \\
$^{10}$ Department of Astronomy, University of California, Berkeley, CA 94720, USA
}

\date{Last updated XXXX XX XX; in original form 2019 September 10}
\pubyear{2019}

\label{firstpage}
\pagerange{\pageref{firstpage}--\pageref{lastpage}}

\maketitle

\begin{abstract}
We present an empirical relation between the cold gas surface density (\Sgas) and the optical extinction (\av) in a sample of 103 galaxies from the Extragalactic Database for Galaxy Evolution (EDGE) survey. This survey provides CARMA interferometric CO observations for 126 galaxies included in the Calar Alto Legacy Integral Field Area (CALIFA) survey. The matched, spatially resolved nature of these data sets allows us to derive the \Sgas-\av relation on global, radial, and kpc (spaxel) scales. We determine \av\ from the Balmer decrement (\ha/\hb). We find that the best fit for this relation is \mbox{$\Sgas~(\msunperpcsq)~\sim~26~\times~ \av ({\rm mag})$}, and that it does not depend on the spatial scale used for the fit. However, the scatter in the fits increases as we probe smaller spatial scales, reflecting the complex relative spatial distributions of stars, gas, and dust. We investigate the \Sgas/\av ratio on radial and spaxel scales as a function of \mbox{$\mathrm{EW(\ha)}$}. We find that at larger values of \mbox{$\mathrm{EW(\ha)}$} (i.e., actively star-forming regions) this ratio tend to converge to the value expected for dust-star mixed geometries ($\sim$ 30 \sigavUnits). On radial scales, we do not find a significant relation between the \Sgas/\av ratio and the ionized gas metallicity. We contrast our estimates of \Sgas using \av with compilations in the literature of the gas fraction on global and radial scales as well as with well known scaling relations such as the radial star-formation law and the \Sgas-\Sstar\  relation. These tests show that optical extinction is a reliable proxy for estimating \Sgas in the absence of direct sub/millimeter observations of the cold gas.
\end{abstract}


\section{Introduction}
\label{sec:intro}

The interstellar medium (ISM) is essential to the understand the structure and evolution of galaxies. The cold ISM is the raw material for the formation of new stars, as is evident from the tight relation between the cold gas surface mass density (\Sgas) and the star formation surface density (\Ssfr), also known as the Star Formation or Kennicutt-Schmidt law \citep{Schmidt_1959,Kennicutt_1998}.  

Radio observations are used to directly trace the mass of the cold gas component in galaxies. The atomic component of the cold gas is observed via the 21 cm line of \hi. Even though the main component of the cold molecular gas in the ISM is \htwo, due to its absence of transitions available at low temperature, emission from CO rotational transitions is used as a proxy for the \htwo\ gas in extragalactic molecular clouds. Large extragalactic single-dish CO and \hi\, surveys have provided estimates of the total molecular and neutral atomic gas mass in galaxies \citep[e.g., FCRAO, COLD GASS surveys;][]{Young_1995, Saintonge_2011} as well as their relevance to galactic processes such as chemical enrichment \citep[e.g.,][]{Peeples_2011}. As noted by \cite{Bolatto_2017} in the EDGE-CALIFA survey presentation paper, spatially resolved information regarding the gas content in galaxies is fundamental to understanding key processes of galactic evolution such as the star formation rate and gas transport. In this regard, spatially resolved maps of CO and \hi\ exist for samples of local galaxies, such as the THINGS/HERACLES surveys \citep{Walter_2008,Leroy_2008}. However, mapping of HI and CO to trace the spatial distribution of neutral and molecular gas is limited and not feasible for large samples of galaxies. This includes the Integral Field Spectroscopy (IFS) surveys \citep[e.g., CALIFA, MaNGA, SAMI;][]{Sanchez_2012,Bundy_2015, Croom_2012} which provide spectral information across their optical extension for a large set of galaxies. Therefore, having an optical tracer for the cold gas distribution for these IFS surveys would be quite useful. 

On the other hand, star-formation is typically embedded or occurs very close to molecular gas complexes which in turn are mixed with dust that attenuates the light from these newly born stars. How the extinction, due to the dust, and the gas correlate to each other could strongly depends on the geometrical arrangement of the stars and the dust \citep[e.g.,][]{Nordon_2013}. Generally, the geometrical arrangements that yield the observed optical extinction (\av) can be classified as a foreground  dust screen obscuring the starlight or a mix of stars and dust. Depending on the physical scale either of these geometries can be assumed \citep[e.g.,][]{Liu_2013,Genzel_2013}. Therefore for a given wavelength, for an obscuring dust screen the extinction would be smaller than for a star-dust mixed geometry. Different studies also consider the impact of tracing \av using different methods in the optical (i.e., emission vs absorption) as well as the spatial distribution of extinction as well as the impact of the diffuse ionized gas in determine \av in galaxies \citep[e.g., ][]{Kreckel_2013,Tomicic_2017}.

Under the assumption of local thermodynamic equilibrium (LTE), it is also expected that the gas column density ($N(H)$) correlates with \av. This relation has been explored extensively within the Milky Way \citep[e.g.,][]{Dickman_1978, Bohlin_1978, Rachford_2009}. More recently the relation between the CO emission (\Ico) and \av has also been explored, resolving clouds of tens of parsecs in size in near extragalactic objects \citep[][]{Lee_2015, Lee_2018}, as well as at kpc scales \citep[e.g., ][]{Boquien_2013}. \cite{Guever_2009} found a tight relation between these two observables using the extinction derived from the Balmer decrement and the hydrogen column density modelling the continuum and the line features from the spectra of 22 supernova remants with  \mbox{$N(H) = (2.21\pm0.09) \times 10^{21} {\rm cm^{-2}} A_V$}. In terms of the total gas surface mass density (\Sgas = \Shtwo + \sighi) this relation translates to 
\begin{equation}
\Sgas = 23 \Big( \frac{A_V}{\rm mag} \Big) (\msunperpcsq).
\label{eq:Guever_09}
\end{equation}
Alternatively, the total gas column density can be obtained assuming an effective gas-to-dust ratio. Assuming a foreground dust screen, \cite{Heiderman_2010} derived $A_V$ using Spitzer infrared SEDs of young stellar objects. Using a constant gas-to-dust relation of \mbox{$N(H) = 1.37 \times 10^{21} {\rm cm^{-2}} A_V$
} with $R_V = 5.5$ \citep{Draine_2003} which includes the helium mass contribution, they found that the gas surface mass density (\Sgas) is given by
\begin{equation}
\Sgas = 15 \Big( \frac{A_V}{\rm mag} \Big) (\msunperpcsq)
\label{eq:Heiderman_2010}
\end{equation}
Dust-star mixed models will yield twice the above values as they account for the dust in the foreground and the background. Optical studies have also provided the relationship between the gas density and the optical extinction. Using the Sloan Digital Sky Survey single-fiber spectroscopic data, \cite{Brinchmann_2013} presented a method to derive the total gas column density from the optical dust attenuation, the dust-to-metal ratio, and the metallicity of the ionized gas for the central region of a large sample of galaxies. In summary, relations between \Sgas and \av can vary depending on the geometry as well as the averaging scale (ranging from pc-size molecular clouds to entire galaxies) \citep[e.g.,][]{Tacchella_2018}. 

Despite the above efforts, there is no systematic study using a homogeneous data set to determine the spatially resolved relation between \Sgas and \av relevant for estimating the gas content in galaxies.  In this regard, the EDGE-CALIFA survey \citep{Bolatto_2017} provides an ideal data set for quantifying the relation between these two parameters on different scales, including galaxy-integrated, radial, and kpc scales. This survey provides a unique observational data set of spatially resolved CO datacubes as well as Integral Field Unit datacubes from the optical CALIFA survey for 126 galaxies in the nearby universe. The main goal of this article is to determine the relation between \Sgas and \av using a large sample of nearby galaxies and to quantify the impact of averaging these properties on different spatial scales.

This paper is organized as follows: in Section~\ref{sec:data} we present the main features of the EDGE-CALIFA survey; in Section~\ref{sec:Analysis} we give a description of the main observables we extract from the different data cubes for each galaxy; in Section~\ref{sec:Results} we present the relation between \Sgas and \av at different spatial scales (from integrated, radial and kpc scales)
; we compare our results with respect to previous Galactic and extragalactic results in Section~\ref{sec:Discussion},  and present the main conclusions of this article in Section~\ref{sec:Conclusions}. 
In this article we assume the following cosmological parameters: $H_0 = 70\ {\rm km\ s^{-1}\ Mpc^{-1}}$, $\Omega_M = 0.27$, $\Omega_\Lambda = 0.73$.

\section{Sample and Data}
\label{sec:data}

The galaxies presented in this study are part of the EDGE-CALIFA survey \citep{Bolatto_2017}. The goal of this survey is to provide  spatially resolved maps of the molecular $^{12}$C$^{16}$O ground rotational transition line ($J = 1 \rightarrow 0$) and at optical wavelengths for 126 galaxies. This surveys enables matched resolution comparison of the properties of the molecular gas with those of other galactic components such as the stellar and ionized gas. In this section we briefly describe the main characteristics of the CALIFA and EDGE surveys as well as our target selection. 

\subsection{The CALIFA survey}
\label{sec:CALIFA}

The CALIFA survey \citep{Sanchez_2012} was designed to acquired spatially resolved spectroscopic information from more than 600 galaxies in the nearby Universe (0.005$< z < $0.03) using the PMAS Integral Field Unit (IFU) instrument \citep{Roth_2005} mounted at the 3.5 m telescope of the Calar Alto Observatory. The main  component of this instrument consists of 331 fibers of 2\farcs7 diameter each, concentrated in a single hexagon bundle covering a field-of-view (FoV) of $74\arcsec\times64\arcsec$, with a filling factor of $\sim$ 60\%\,.  Three-point dithering allows a full coverage of the FoV. The nominal resolution of this instrument is $\lambda$/$\Delta\lambda$\,$\sim$\,850 at $\sim$5000\AA\, with a nominal wavelength range from  3745 to 7300\AA. Besides the restriction in redshift, most CALIFA galaxies are expected to match the instrument FoV. Their isophotal diameters in the SDSS $r$-band are in the range 45 $\lesssim D_{25}\lesssim$\,80 arcsec \citep{Walcher_2014}. The data reduction is performed by a pipeline designed specifically for the CALIFA survey. The final data cube for each galaxy consists of more than 5000 spectra with a sampling of 1$\times$1 arcsec$^2$ per spaxel. The detailed reduction process is described in \cite{Sanchez_2012}, and improvements on this pipeline as well as extensions to the original sample (reaching a total of 834 galaxies) are presented by \cite{Husemann_2013,GarciaBenito_2015,Sanchez_2016}. 

\subsection{The EDGE survey}
\label{sec:EDGE}
\begin{figure*}
\includegraphics[width=\linewidth]{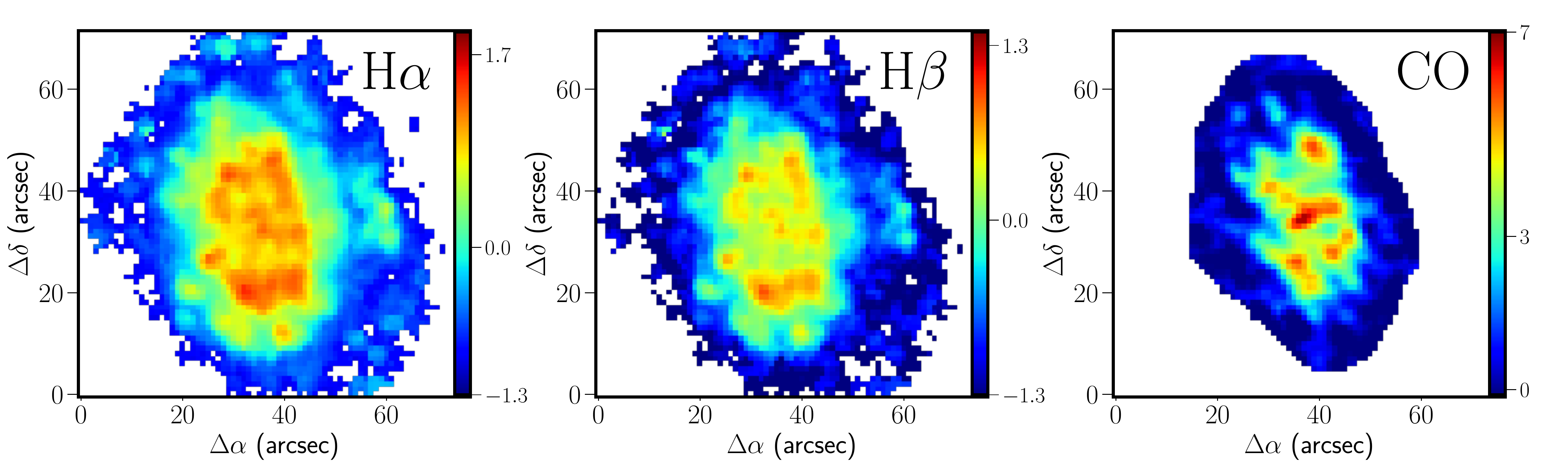}
\caption{Example of the maps used for this study. Left and middle panels correspond to maps of the logarithm of the flux from the \ha and \hb emission lines for the galaxy NGC~5633. Both in units of \mbox{$10^{-16}\mathrm{erg\,s^{-1}\,cm^{-2}}$}. The right panel corresponds to the CO intensity in units of \Kkmpers\, for the same object. The CO intensity maps has been centered and regrided to have the same spatial resolution as the CALIFA datacubes \citep[1 arcsec,][]{Utomo_2017}. To compare the \av derived from the Balmer decriment with the \Sgas derived from the CO intensity, our selection considers only those regions where both \ha and CO emission are detected.}
\label{fig:Maps}
\end{figure*}%
The EDGE survey obtained millimeter-wave interferometric observations for a subsample of galaxies selected from the CALIFA survey.  These observations were carried  out at the Combined Array for Millimeterwave Astronomy \citep[CARMA, ][]{Bock_2006}.  The EDGE survey provides the first major effort to combine CO data with IFS optical data. 

We present a brief description of the survey here. See \cite{Bolatto_2017} for a detailed description. Observations took place between November 2014 and April 2015. Galaxies were observed using half-beam-spaced seven-point hexagonal mosaics to optimize extended flux recovery over the central 1 arcmin region and yielding a half-power field-of-view of radius $\sim$ 50\arcsec.  177 CALIFA galaxies were observed in the E-array configuration (each with typically 40 min of integration time and $\sim$ 8 arcsec resolution). From these galaxies, 126 with detections or possible detections of CO emission were selected for an additional $\sim$ 3.5 hr integration, this time in the more extended D-array, yielding $\sim$ 4\arcsec resolution, equivalent to $\sim$ 1-2 kpc at the typical distances of galaxies in the survey. To maximize observing efficiency,  galaxies were observed in different groups according to their redshift. Each of these observation bins used a fixed tuning and correlation set-up. The correlator covers five 250-MHz windows spanning the velocity range where emission is expected to occur for each of the targets. The final maps combined the E and D array observations resulting in  velocity resolution of 20 km s$^{-1}$ with a typical angular resolution of 4.5\arcsec and typical rms sensitivity of 30 mK at the velocity resolution. Assuming a constant CO-to-\htwo\, conversion factor, the survey is sensitive to an \htwo\, surface mass density of  $\sim\,4-110\,\,\mathrm{M_{\odot}\,\,pc^{-2}}$ (averaged over a $\sim$ 1.5 kpc scale). The data cubes are smoothed and then masked in order to distinguish CO signal from noise and to reach higher signal to noise \citep[see more details in][]{Bolatto_2017}. Projects conducted with EDGE data to date include in-depth studies of the impact of morphology and kinematics on the molecular depletion time \citep{Utomo_2017, Colombo_2018}, as well as dynamical comparisons of the cold and ionized gas \citep{Levy_2018} and dynamical modelling of different baryonic components \citep{Leung_2018}.

\section{Analysis}
\label{sec:Analysis}

\subsection{CALIFA and EDGE Spatially Resolved Properties}
\label{sec:maps}

We used the IFU analysis pipeline \texttt{PIPE3D} \citep{Sanchez_2015} in order to extract physical parameters from the CALIFA data cubes. A full description of how the pipeline extracts two-dimensional physical properties from data cubes is described in \cite{Sanchez_2016}. The pipeline extracts properties from a large variety of emission lines included in the covered wavelength range such as integrated flux, equivalent width, line-of-sight velocity, and velocity dispersion. In particular, for this study we use fluxes from the emission lines H$\alpha$, H$\beta$, [{\rm O{\small III}}]$\lambda$5007\AA\,, and [{\rm N{\small II}}]$\lambda$6583\AA\, as well as the H$\alpha$ equivalent width (EW(H$\alpha$)). The pipeline also provide a myriad of information regarding the stellar component including the stellar mass density \Sstar \citep[see details in][]{Sanchez_2016}. For the resolved analysis, we use the photometric axis ratio for each galaxy \citep{Walcher_2014} to correct the surface densities for inclination effects \citep{BB2016}. 

To obtain the velocity-integrated CO surface brightness maps (\Ico), the smoothed and masked  EDGE data cubes are integrated along the velocity axis. The uncertainties are also derived from the integrated noise data cubes. We convert \Ico into molecular gas surface density  (\Shtwo) via a constant CO-to-\htwo\, conversion factor (\aco) of 4.4 \acounits \citep{Bolatto_2008}, including the mass contribution from helium.

In order to compare properties from the two data sets on a spaxel by spaxel basis, we follow a similar procedure to that described in \cite{Utomo_2017}. The EDGE maps are centered and regrided to match the spaxel size and grid of the CALIFA maps by using the \texttt{MIRIAD} task \texttt{regrid}. The CALIFA maps are also convolved to match the resolution of EDGE maps. In Fig.~\ref{fig:Maps} we show an example of the emission maps used for this analysis. 

From the emission line flux maps, we derive the dust attenuation for the \ha emission line ($A(\mathrm{\ha})$) following \cite{Catalan-Torrecilla_2015}. Assuming an extinction-free \ha / H$_{\beta}$ flux ratio of 2.86 \citep{Osterbrock_1989} and $R_V$ = 3.1 \citep{Cardelli_1989},
\citep{Cardelli_1989},
\begin{equation}
A(\ha) = \frac{K_{\rm H\alpha}}{-0.4(K_{\rm H\alpha} - K_{\rm H\beta})} \times \log\left(\frac{F_{\rm H\alpha}/F_{\rm H\beta}}{2.86}\right)
\label{eq:Abalmer}
\end{equation}
where $F_{\rm H\alpha} / F_{\mathrm{H}\beta}$ is the flux ratio between these Balmer lines, and $K_{\mathrm{H}\alpha}$ = 2.53 and $K_{\mathrm{H}\beta}$ = 3.61 are the extinction coefficients for the Galactic extinction curve from \cite{Cardelli_1989}. The values of $K_{\mathrm{H}\alpha}$ and $K_{\mathrm{H}\beta}$ are similar for both \cite{Cardelli_1989} and \cite{Calzetti_2000} extinction curves and dust-to-star geometries. Assuming an extinction curve of $\mathrm{R_V}$ = 3.1 \citep{Cardelli_1989}, the optical extinction is given by
\begin{equation}
\mathrm{A_V} =  A(\mathrm{H}\alpha) / 0.817. 
\end{equation}

We use the luminosity of the \ha emission line to derive the integrated and spatially resolved star formation rates (SFR) presented in this study. We transform this luminosity to SFR following the relation presented in \citet{Kennicutt_1998_SFR} which assumes a Salpeter Initial Mass Function and solar metallicity
\begin{equation}
{\rm SFR\,(M_{\odot}\,yr^{-1})}=8\times10^{-42} L{\rm(H\alpha)} ({\rm erg\,s^{-1}})
\label{eq:SFR_Ha}
\end{equation}
where $L{\rm(H\alpha)}$ is the luminosity of the \ha emission line obtained using the extinction-corrected integrated flux ($F_{\rm corr}{\rm(H\alpha)}$) via
\begin{equation}
L{\rm(H\alpha)} ({\rm erg\,s^{-1}}) =  1\times10^{-16} \times 4\pi D^2 F_{\rm corr}{\rm(H\alpha)} 
\end{equation}
where $D$ is the luminosity distance for each galaxy in units of ${\rm cm}$ and $F_{\rm corr}{\rm(H\alpha)}$ has units of $10^{-16} {\rm erg\,s^{-1}\,cm^2}$. In Sec.~\ref{sec:Global} we used the \ha equivalent width  measured at an effective radius of one ($\mathrm{EW (\ha )_{eff}}$); this value was derived for all the CALIFA galaxies in \cite{Sanchez_2018}. Finally, we derived the metallicity for each of the selected spaxels in the EDGE-CALIFA galaxies. Here we use the metallicity from the O3N2 empirical calibrator \citep{Marino_2013}. A full description of the derivation of  metallicity for CALIFA galaxies using this calibrator is presented in \cite{Sanchez_2017}.

\section{Results}
\label{sec:Results}
Due to the spatially-resolved nature of the EDGE data cubes,  we can study the correlation between molecular gas surface density \Sgas and  extinction \av in three different spatial regimes: global or galaxy-integrated measures (Sec.~\ref{sec:Global}), radial bins (Sec.~\ref{sec:Radial}), and spaxel-by-spaxel comparisons (Sec.~\ref{sec:Spaxels}). 

\subsection{Global Properties}
\label{sec:Global}
\begin{figure}
\includegraphics[width=\linewidth]{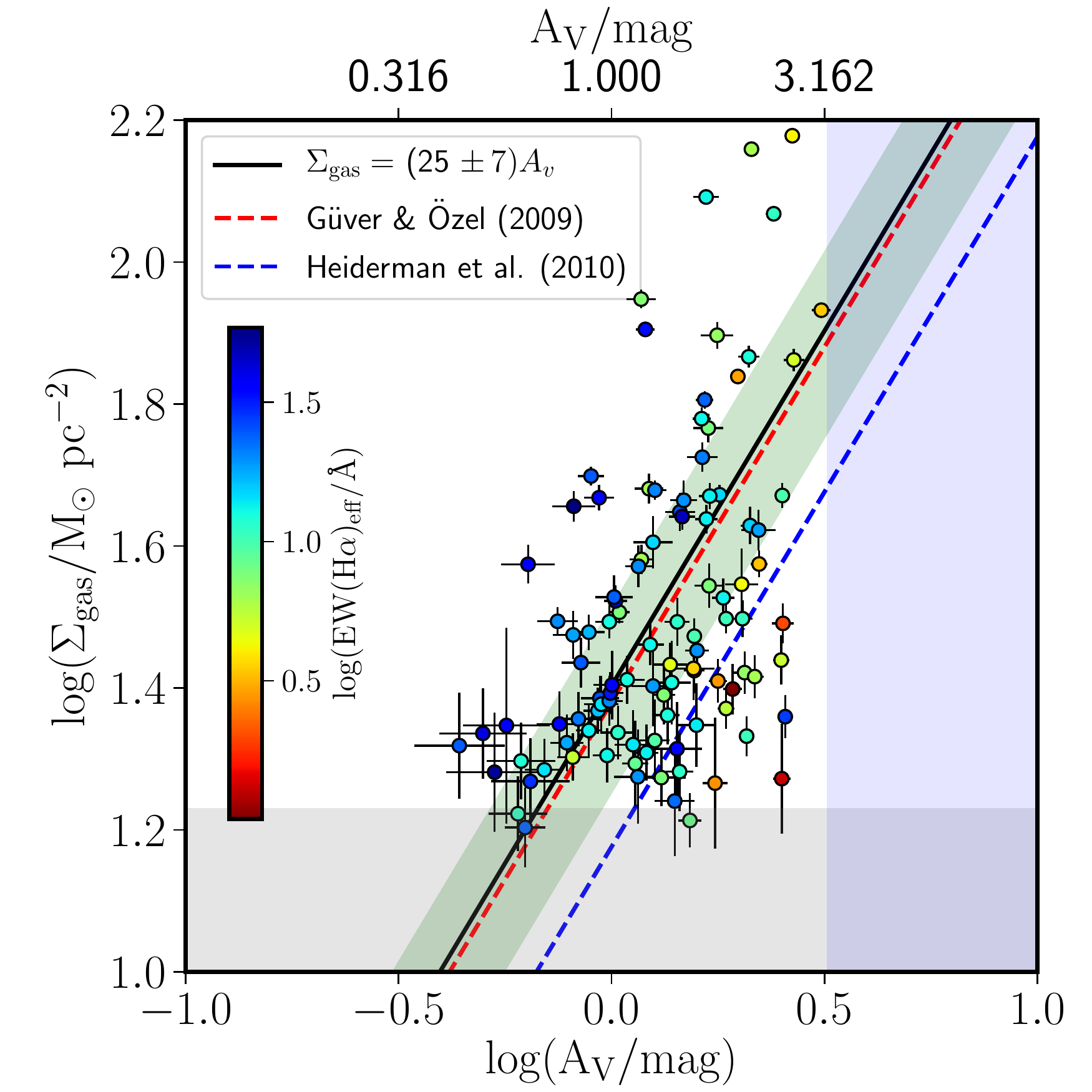}
\caption{Comparison of the galaxy-averaged gas mass surface density and optical extinction for the selected EDGE galaxies (see Sec.~\ref{sec:Global} for details). Symbols are color-coded by the \ha equivalent width at the effective radius ($\mathrm{EW (\ha )_{eff}}$). Horizontal-gray and vertical-blue shaded areas represent the detection limits for \Sgas and \av, respectively. The black line represents the best fitted line of the form $\Sgas = b\, \mathrm{ A_V}$, with $b = 25 \, \mathrm{M_{\odot} \,pc^{-2} \,mag^{-1}} $; the green shaded area reflects the uncertainty from CO selection effects (see text for details). The red and blue-dashed lines represent the \Sgas-\av relation derived from observations \citep[cf.~Eq.(\ref{eq:Guever_09}),][]{Guever_2009} and  a dust-screen model \citep[cf.~Eq.~(\ref{eq:Heiderman_2010}),][]{Heiderman_2010}. Galaxies with larger  $\mathrm{EW (\ha )_{eff}}$ tend to have a similar \mbox{\Sgas-\av} relation as the one derived from observations.} 
\label{fig:Int_Av_SCO}
\end{figure}

Prior to obtaining the integrated properties from the optical and millimeter maps, we select those regions with non-negative flux in both the CO and \ha emission maps. Although this selection criterion may consider spaxels with relatively low signal to noise ratio, or even below the detection limit, it does provide a good compromise between reliable detections and upper limits. Below we consider the effect of different CO flux selection thresholds in estimating the best relation between \Sgas and \av. It also provides a  reliable way to mimic the integrated properties derived using a single-dish in the millimeter regime or narrow filters in the optical.  

For each of the selected spaxels in each galaxy we derived the extinction from the Balmer decrement as described in Sec.~\ref{sec:Analysis}. In those cases where $\ha / \hb < 2.86$ with CO flux larger than zero,  we assume a median extinction derived from those spaxels in the galaxy with  $\ha / \hb > 2.86$. Following \cite{Sanchez_2018},  we derive the extinction for each galaxy as the median value obtained for all the spaxels that satisfy the above selection criteria. We note that extinction derived using this method is quite similar to the one derived from the ratio of the integrated \ha and \hb fluxes in each galaxy. For the CO datacubes, we derive the total mass of the molecular hydrogen in each galaxy by adding up the mass contributions from individual spaxels assuming a conversion factor of \aco~=~$4.4~\times~10^{20}~$\acounits. To derive the total molecular mass surface density (\Shtwo), we divide the total mass by the effective area, which is the area of a single spaxel (in pc$^2$) times the number of selected spaxels in each galaxy. Finally to compare \av to the total hydrogen gas density (\Sgas = \Shtwo + \sighi), since resolved HI maps are not available for most galaxies we use a fiducial HI surface density of \sighi = 6 \msunperpcsq. This deprojected surface density is a typical value in disc galaxies where \htwo\ is detected \citep[]{Bigiel_2008}.  To provide robust estimates of the relation, we exclude from our sample galaxies with values of \av $<$ 0.2. That is, we exclude those galaxies/regions where CO emission likely drops rapidly due to photo dissociation of the CO molecules \citep[]{van_Dishoeck_1988}. The final sample consists of 103 galaxies out of the total sample of EDGE galaxies (126 targets).

In Fig.~\ref{fig:Int_Av_SCO} we plot the galaxy-averaged values of \Sgas vs \av for the sample of selected EDGE galaxies. As noted above, the fiducial values of \Sgas depend on the assumed thereshold to detect CO flux. For our particular selection (CO flux larger than zero) we represent this limit in Fig.~\ref{fig:Int_Av_SCO} with  horizontal-gray shading. The upper value of the shading represents the sum of the assumed value of \sighi and the 3$\sigma$ brightness sensitivity of the EDGE-CALIFA survey \citep[\Shtwo $\sim 11 \msunperpcsq$,][]{Bolatto_2008}. On the other hand, for our data we are not able to estimate \av larger than 3 mag due to the impossibility of measure \hb\ at such large extinction. We indicate this limit with vertical blue shading. Despite the scatter, \Sgas increases with \av.  

We also overplot the expected relations presented in Eqs.~\ref{eq:Guever_09} and \ref{eq:Heiderman_2010} (red and blue dashed lines, respectively). We note that in general the integrated \Sgas and \av follows these two lines. Following this relation, we fit this data set using a linear relation of the form 
\begin{equation}
\Sgas = b \av,
\label{eq:SgasAv}
\end{equation}
with $b = 25\, \mathrm{M_{\odot} \,pc^{-2} \,mag^{-1}}$ determined as the best fitting value (see black line in  Fig.\ref{fig:Int_Av_SCO}). We note than this value is more similar to the one determined from observations by \cite{Guever_2009} than the expected  value from a dust-screen geometry obtained by \cite{Heiderman_2010}. When we repeat the above analysis selecting only low-inclination galaxies (i.e.,  $i < 65^{\circ}$), we obtain a similar value for $b$, $b = 23 \, \mathrm{M_{\odot} \,pc^{-2} \,mag^{-1}} $. To account for the fact that different CO flux selection can lead to different estimates of the best fit \Sgas-\av relation, the green-shaded area in  Fig.~\ref{fig:Int_Av_SCO} represents the range of fits using two different selection criteria. By using a conservative selection criteria with CO detection brighter than 2$\sigma$ we find the upper envelope of this area (i.e.,  $b = 29 \, \mathrm{M_{\odot} \,pc^{-2} \,mag^{-1}} $). This value is significantly larger than the one derived by \cite{Guever_2009}. However, it is similar to the expected value for a dust-star mix geometry, where $b$ is expected to be close to twice the value derived for the dust-screen geometry \citep{Nordon_2013}. On the other hand, if we loosen this constraint and select all the CO flux, we find values very similar to those reported by \cite{Heiderman_2010}, $b = 18 \, \mathrm{M_{\odot} \,pc^{-2} \,mag^{-1}}$.

The equivalent width of the \ha\, emission line ($\mathrm{EW(\ha)}$) has been used extensively to quantify the nature of the ionization source responsible for the emission from the ionized gas in extragalactic objects \citep[e.g., ][]{CidFernandes_2010, BB2016}. Large values of $\mathrm{EW(\ha)}$ strongly correlate with underlying young stellar population and thus recent star formation \cite[e.g.,][]{Sanchez_2015} whereas low-values are attributed to ionization due to old-stellar population or any other process of ionization such as underlying old stellar population, shocks, etc  \citep[e.g.,][with the exception of ionizition due to Active Galactic Nuclei]{Lacerda_2018}. In Fig.~\ref{fig:Int_Av_SCO} we color-coded the data points according to the $\mathrm{EW(\ha)}$ measured at the effective radius for each galaxy ($\mathrm{EW(\ha )_{eff}}$). Altough, we find that the median value of the ratio \Sgas/\av increases as we select galaxies with larger $\mathrm{EW(\ha )_{eff}}$, the typical value of this ratio for this sample is similar to the best fit of Eq.~\ref{eq:SgasAv} ($\Sgas/\av \sim 25 \,\mathrm{M_{\odot} \,pc^{-2} \,mag^{-1}} $). Selecting galaxies with $\mathrm{EW(\ha )_{eff}} > 6 {\rm \AA}$ (the typical value used to segregate star-forming galaxies) we find a similar  median ratio to the one derived using the entire sample. However, when selecting galaxies with $\mathrm{EW(\ha )_{eff}} > 20 {\rm \AA}$ we find that the median value of the \Sgas/\av ratio agrees with the value expected for a dust-star mixed geometry ($\Sgas/\av \sim 30 \,\mathrm{M_{\odot} \,pc^{-2} \,mag^{-1}} $). These results suggest that for actively star-forming galaxies (i.e., galaxies with large $\mathrm{EW(\ha)}$ values), the relation between the gas density and the optical extinction is similar to a dust-star mix configuration rather than the commonly assumed dust-screen geometry. 

In summary, these results suggest that the galaxy-averaged \Sgas can be derived from the galaxy-averaged \av. However, due to the limits inherent on the  detection sensitivity for both gas (at low values of \Sgas, CO detection) and extinction (at large values of \av, \hb detection) the linear relation expected from these two values is affected. Even more, we also note that galaxies without current star formation (i.e., galaxies with small  $\mathrm{EW(\ha)}$ values) contribute to the scatter of the global \Sgas-\av relation. 

\begin{figure}
\includegraphics[width=\linewidth]{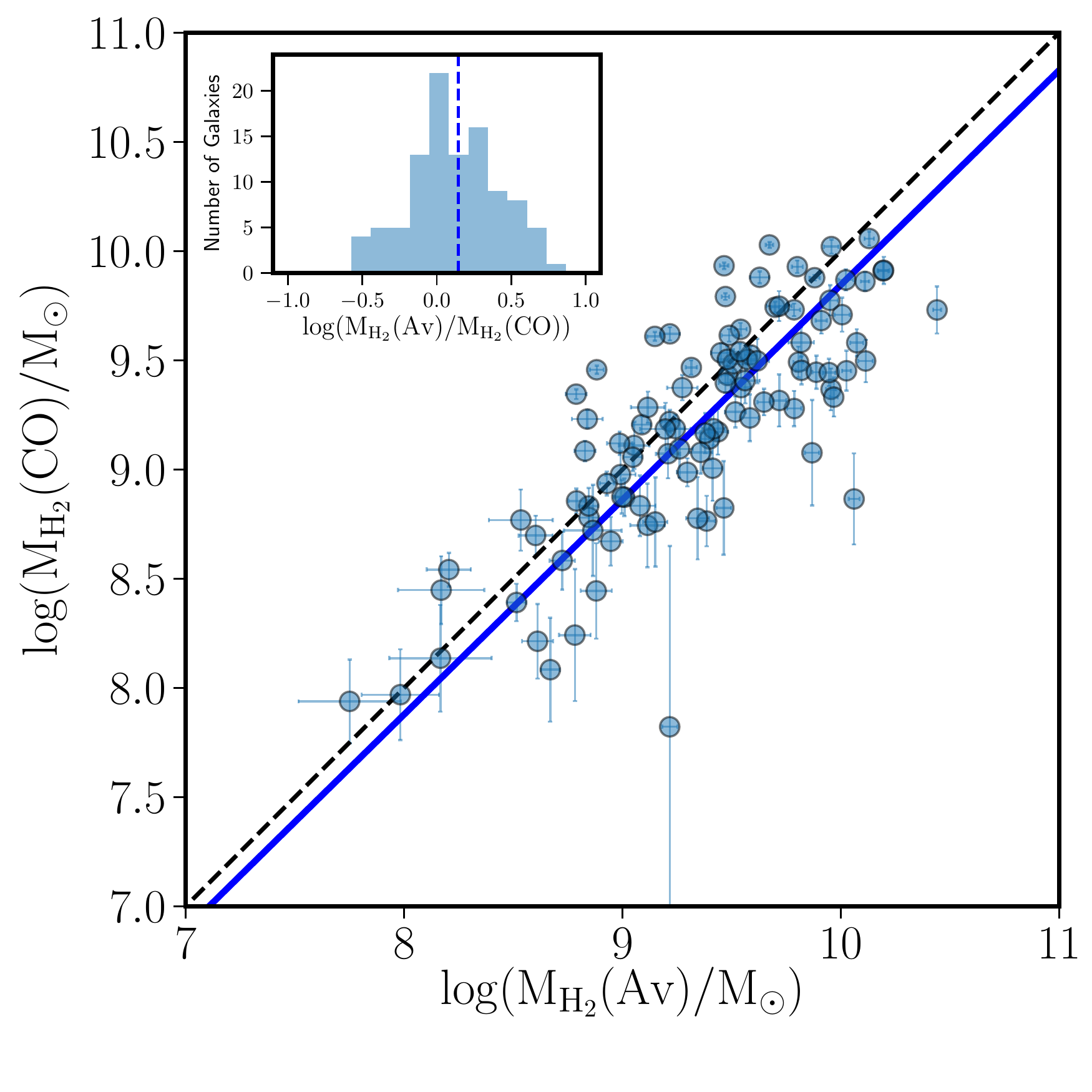}
\caption{Comparison of the total molecular gas mass for EDGE galaxies derived from direct CO observations \citep[${\rm M_{H_{2}}(CO)}$, ][]{Bolatto_2017} and the molecular mass derived  using \av as proxy  from this work (${\rm M_{H_{2}}(\av)}$). The best fit (blue line) is similar to the expected one-to-one relation (black-dashed line). The inset shows the distribution of the ratio between the molecular mass obtained from each of the two methods; the median value of this ratio is close to one (vertical blue-dashed line). \av is a good indicator of the molecular gas mass content of a galaxy.} 
\label{fig:Mass_CO_Av}
\end{figure}

The above relation also allows us to compare the estimate of the total galaxy molecular gas mass from the integrated CO flux with the one using extinction as a proxy for gas mass surface density. To derive molecular gas mass from \av in each galaxy, first we use the relation at spaxel scales to transform \av to \Sgas (see details in Sec.~\ref{sec:Spaxels}). Next, we derived the mass in each spaxel multiplying  \Sgas times the spaxel area. Then, we integrate the individual contributions of spaxels to derive the total gas mass for each galaxy. Finally, to obtain the total molecular mass from the extinction (${\rm M_{H_{2}} (\av)}$),  we subtract the contribution of  \hi\ to the total gas mass. 

In Fig.~\ref{fig:Mass_CO_Av} we plot the total mass of ${\rm M_{H_{2}}}$ for each EDGE-CALIFA galaxy using CO (with ${\rm M_{H_{2}} (CO)}$ from \citealt{Bolatto_2017}) vs the total mass of ${\rm M_{H_{2}}}$ estimated as described above using \av (i.e., using \Sgas = 25 (\sigavUnits) \av). Despite the difference in methods for deriving ${\rm M_{H_{2}}}$, we find that most galaxies lie near the one-to-one line (dashed-line in Fig.~\ref{fig:Mass_CO_Av}). 

To further quantify  possible differences in the two methods for estimating molecular gas mass, in the inset of Fig.~\ref{fig:Mass_CO_Av} we show the distribution of the ratio of the molecular mass obtained from each method. The distribution is quite tight and close to unity; indeed the average gas mass ratio is 1.1 and its standard deviation is similar to the deviation of the relative errors of the gas masses ($\sim$ 0.3 dex).  The best fit using an orthogonal distance regresion (ODR) between these two masses (blue line in Fig.~\ref{fig:Mass_CO_Av}) is close to a linear relation with a best fitted intercept and slope in logarithm scale of $(0.1\pm0.5) \log (\mathrm{M_{\odot}})$ and $0.98\pm0.05$, respectively.  

\subsection{Radial binned properties}
\label{sec:Radial}
\begin{figure}
\includegraphics[width=\linewidth]{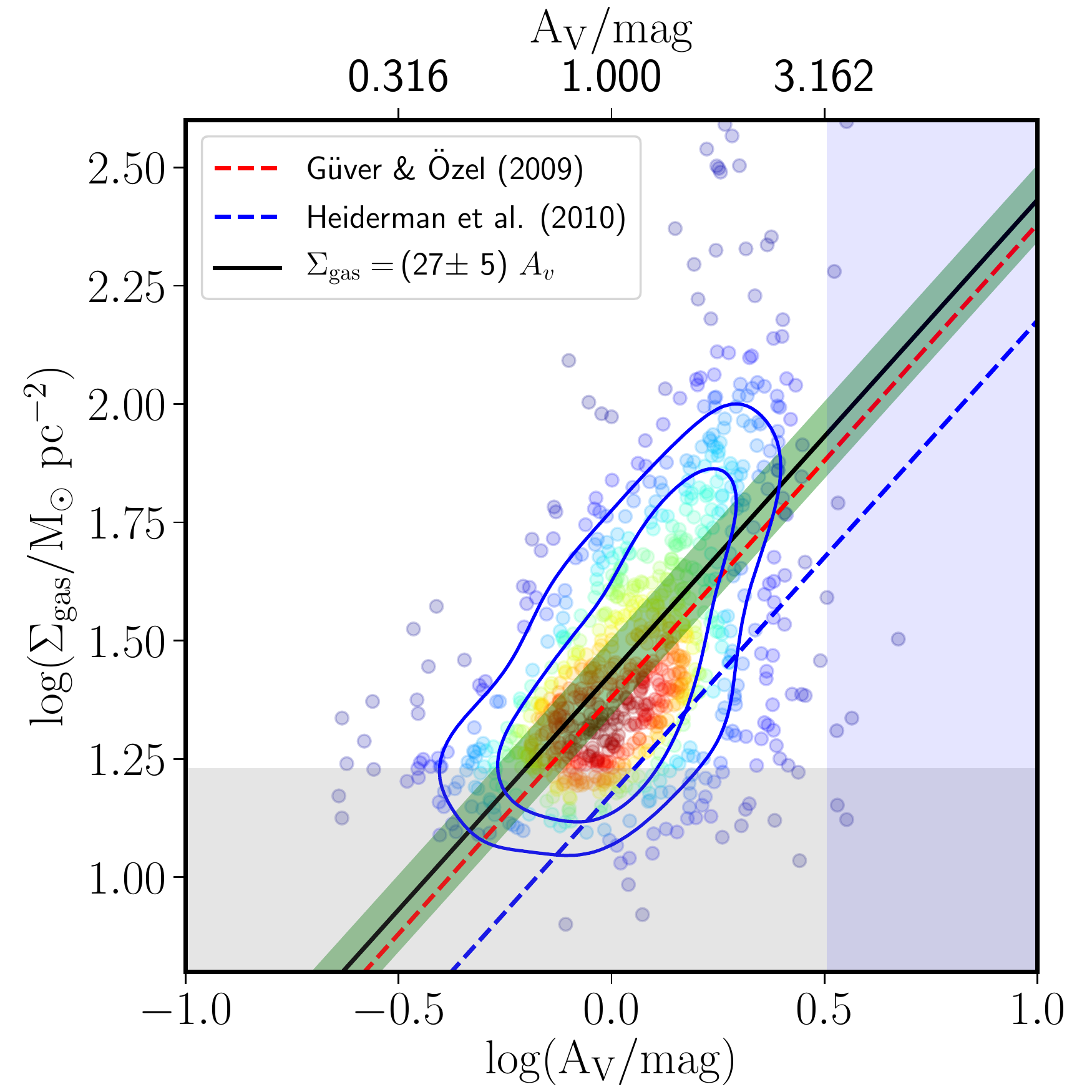}
\caption{\Sgas as a function of \av derived in annuli of 0.2 \Reff width. Data points are color-coded according to their density distribution. The blue outer and inner contours enclose 80\% and 60\% of the sample, respectively. As in Fig.~\ref{fig:Int_Av_SCO}, horizontal-gray and vertical-blue shaded regions represent the detection limits for \Sgas and \av, respectively. The black line and green-shaded  area represent the best fit line and $\pm~1\sigma$ region of the form \Sgas = $b$ \av , with $b~=~(27~\pm~5)\,\mathrm{M_{\odot} \,pc^{-2} \,mag^{-1}} $. Red and blue dashed lines represent Eqs.~\ref{eq:Guever_09} and \ref{eq:Heiderman_2010}, respectively. At these spatial scales, \av is a reliable indicator of the gas content in galaxies.}
\label{fig:Rad_Av_SCO}
\end{figure}

The spatially resolved data allows us to study the relationship between \Sgas and \av in radial bins in our sample of extragalactic objects. We averaged the spatial distribution of each observable in annuli of 0.2 \Reff\, width each out to a radius of 2.5 \Reff. To account for the effects of inclination, we use  the ellipticities and position angles provided for the CALIFA galaxies \citep{Walcher_2014}. For our analysis, in each galaxy we select those annuli for which at least 15 per cent of the spaxels included in the annulus have reliable CO detections ($> 1\sigma$). To derive the extinction for each of the selected annulus, we compute the average \ha and \hb fluxes within it and estimate the extinction from their ratio, as described in Sec.~\ref{sec:Analysis}. We derive the integrated CO flux in each selected annulus using all the unmasked CO spaxels to determine its corresponding \Shtwo using a similar procedure described above (Sec.~\ref{sec:Analysis}). Similar to our procedure for  galaxy-integrated properties, we assume a constant density for the neutral gas component. Thus in each annulus \Sgas = \Shtwo + \sighi, with \sighi = 6 \msunperpcsq. As result,  we are able to measure \Sgas and \av in 893 annuli from 93 EDGE-CALIFA galaxies. 

In Fig.~\ref{fig:Rad_Av_SCO} we plot the comparison between \Sgas and \av derived for each annulus. \Sgas increases with \av. As we describe in Sec.~\ref{sec:Global}, here we are also limited by our ability to detect CO and \hb emission line flux in these annuli (horizontal-gray and vertical-blue shades in Fig.~\ref{fig:Rad_Av_SCO}). We note that \Sgas and \av follow a similar relation as the one derived at  galaxy-integrated scales (see Fig.~\ref{fig:Int_Av_SCO}).

We overplot blue and red-dashed lines representing the relations from \cite{Heiderman_2010} and \cite{Guever_2009}, respectively. We note that a significant fraction of points have values that lies between these two lines. Using orthogonal distance regression (ODR), we fit Eq.~\ref{eq:SgasAv} to 80 per cent of the sample (see data points enclosed in the blue contour in Fig.~\ref{fig:Rad_Av_SCO}). This fit results in $b~=~(27~\pm~5) \, \mathrm{M_{\odot} \,pc^{-2} \,mag^{-1}}$. We derived the uncertainty in $b$ by fitting this parameter in a  Monte Carlo simulation in which we allow the following parameters to vary: ({\it i}) \sighi = [3,8] \msunperpcsq; these values are typically obtained from \sighi radial profiles observed in nearby galaxies \citep{Bigiel_2008}; ({\it ii}) the choice of CO brightness threshold  was varied from 0.5 to 2 $\sigma$; and ({\it iii})  the minimum percentage of spaxels per annulus with CO flux was varied from 15 to 45 percent. 

The best fit radial \Sgas-\av relation is in agreement with the estimates from \cite{Guever_2009} rather than the values expected from a dust-screen model \citep{Heiderman_2010}. On the other hand, the scatter of this relation derived for radial bins is larger than the one derived using integrated properties (0.73 dex and 0.55 dex, respectively). Despite the large differences in physical scales, these results indicate that as we measure these two observables at smaller angular scales, the scatter of the \Sgas-\av relation increases, suggesting a complex structure of geometries between the stars and the dust/molecular gas. In Sec.\ref{sec:EW} we explore the impact of the $\mathrm{EW(\ha)}$ on the \Sgas-\av relation. In particular, we will quantify whether this parameter can separate the wide range of geometries  observed at radial scales. Despite the scatter, we consider that the average \av measured in radial bins is a good indicator of the gas density of the ISM in nearby galaxies.

\subsection{Spaxel-by-spaxel Properties}
\label{sec:Spaxels}
\begin{figure}
\includegraphics[width=\linewidth]{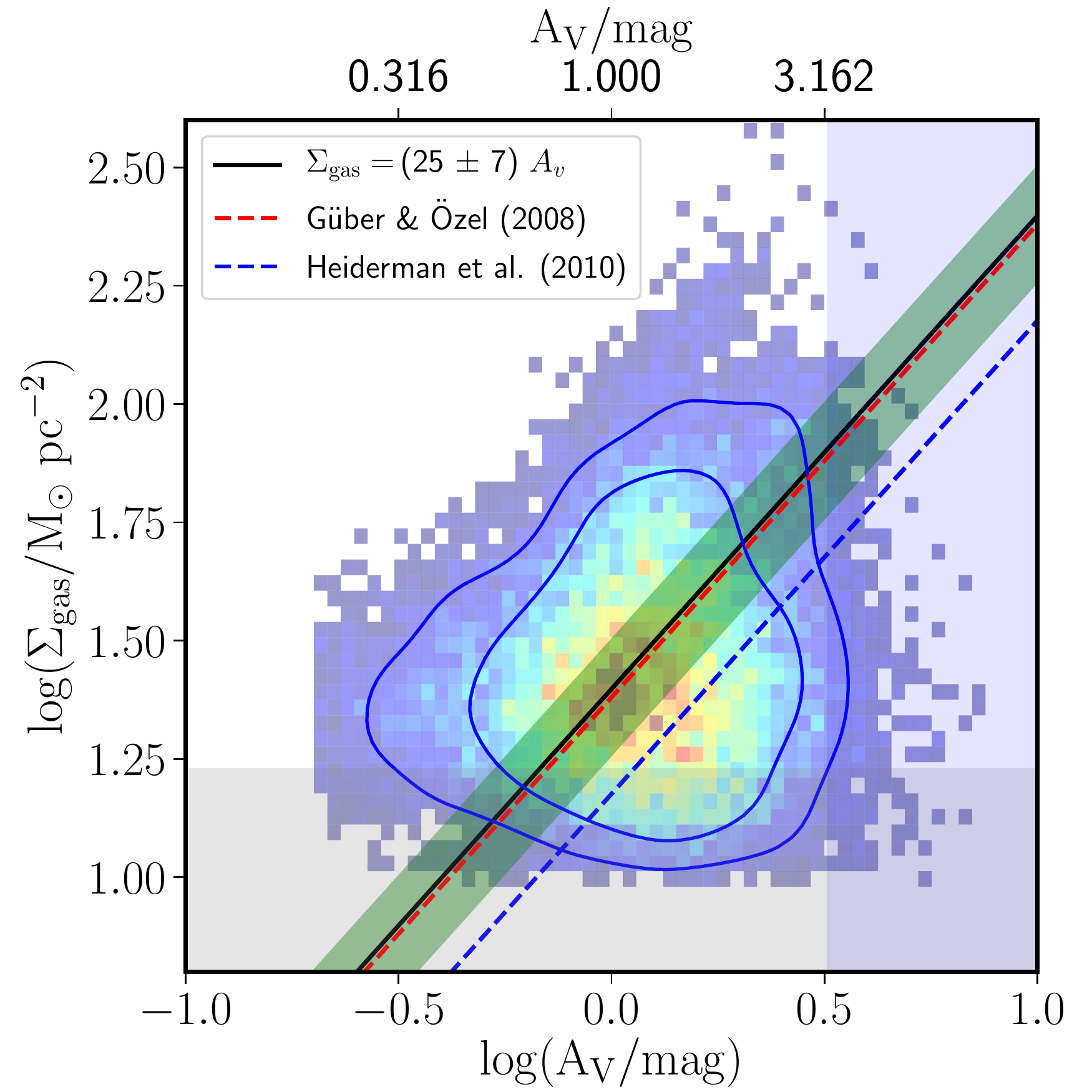}
\caption{Distribution of \Sgas versus \av at spaxel scales. As in  previous figure, the distribution is color-coded to represent its density distribution.  The outer and inner blue contours enclose 80\% and 60\% of the sample, respectively. The black line and green-shaded areas represent the best fit and uncertainties using Eq.~\ref{eq:SgasAv}, with $b~=~(25~\pm~7) \, \mathrm{M_{\odot} \,pc^{-2} \,mag^{-1}} $.  The blue and red-dashed lines represent the relations from \protect\cite{Heiderman_2010} and \protect\cite{Guever_2009}, respectively. It is evident that as we probe smaller physical scales, the \Sgas-\av relation shows larger scatter.}
\label{fig:spaxel_Av_SCO}
\end{figure}

In Fig.~\ref{fig:spaxel_Av_SCO} we plot the relationship of  \Sgas and \av derived on a spaxel-by-spaxel basis for the galaxies. For this analysis we select spaxels with reliable CO detections ($>$1$\sigma$) and reliable measurements of extinction. To determine \Shtwo and \av we follow Sec.~\ref{sec:Analysis}. We correct \Shtwo for inclination following \cite{BB2016}. This selection criteria allow us to measure this properties in 34705 spaxels located in 103 EDGE-CALIFA galaxies. 

As in Sec~\ref{sec:Radial}, we use \Sgas = \Shtwo + \sighi, with \sighi = 6 \msunperpcsq. Although, the observed distribution of spaxels in the \Sgas-\av plane shows a large dispersion in comparison to the radial distribution (see Fig.~\ref{fig:Rad_Av_SCO}), the trend of  \Sgas and \av is similar to those derived at radial and integrated scales. This also implies that bulk of the spaxels have \Sgas and \av that lies between the relations found in the literature (blue and red dashed lines in  Fig.~\ref{fig:Rad_Av_SCO}). We note that low values of \av ($<$ 0.5) tend to have a relatively constant value of \Sgas ($\sim$~20~\msunperpcsq). A possible explanation for this tail at low \av is the CO detection limit of ($\sim$~11~\msunperpcsq) as well as our assumption of a constant \sighi which at these scales may not be true for all the galaxies where we detect CO. We also have a lower limit in determining \av due to the detection of the \hb emission line. These limitations are represented in Fig.~\ref{fig:spaxel_Av_SCO} by the horizontal and vertical dashed areas. We fit Eq.~\ref{eq:SgasAv} to 80 per cent of the data (based on the data density distribution.) using ODR. We find the best fitted value to be $b~=~(25~\pm~7) \, \mathrm{M_{\odot} \,pc^{-2} \,mag^{-1}}$. The uncertainty for this parameter reflects the previous assumptions we made regarding our ability to detect CO and the value of \sighi. We use a monte carlo simulation of 1000 random values of the limit in CO detection (0.5$<\sigma<$2) and \sighi = [3,8] \msunperpcsq; which, as we mention above, is the typical range observed in local galaxies. 

Despite the scatter, we note that the best fit value at spaxel scales is closer to the relation derived by \cite{Guever_2009}. Comparing to the relations derived at other spatial scales (radial and global), this best fit is similar to the other physical scales. These results agree with the previous results at radial scales, meaning that the smaller the physical scale we probe, the larger the scatter we find in the \Sgas-\av relation. This large scatter  suggests the complex interplay between the distribution of the gas/dust and the location of the stars at kpc scales. As for the radial scales, in  Sec.\ref{sec:EW} we explore the impact of the $\mathrm{EW(\ha)}$ on the \Sgas-\av relation at spaxel scales. In particular, we will quantify whether this parameter can separate the wide range of geometries observed at radial scales. 

\subsection{The impact of $\mathrm{EW(\ha)}$ on the resolved \Sgas-\av relation}
\label{sec:EW}
In Sec.~\ref{sec:Global} we note that for galaxies with larger values of $\mathrm{EW(\ha )_{eff}}$ (i.e., actively star-forming galaxies) the \Sgas-\av ratio is approximately twice larger than the value expected for an obscuring dust geometry \citep[e.g,][]{Heiderman_2010}. This suggests that for star-forming galaxies the dust and  gas are mixed with stars resulting in an even-mix geometry. Following this study, we explore in this section the impact of the $\mathrm{EW(\ha)}$ on the \mbox{\Sgas-\av} ratio at radial and spaxel scales. 

\begin{figure}
\includegraphics[width=0.98\linewidth]{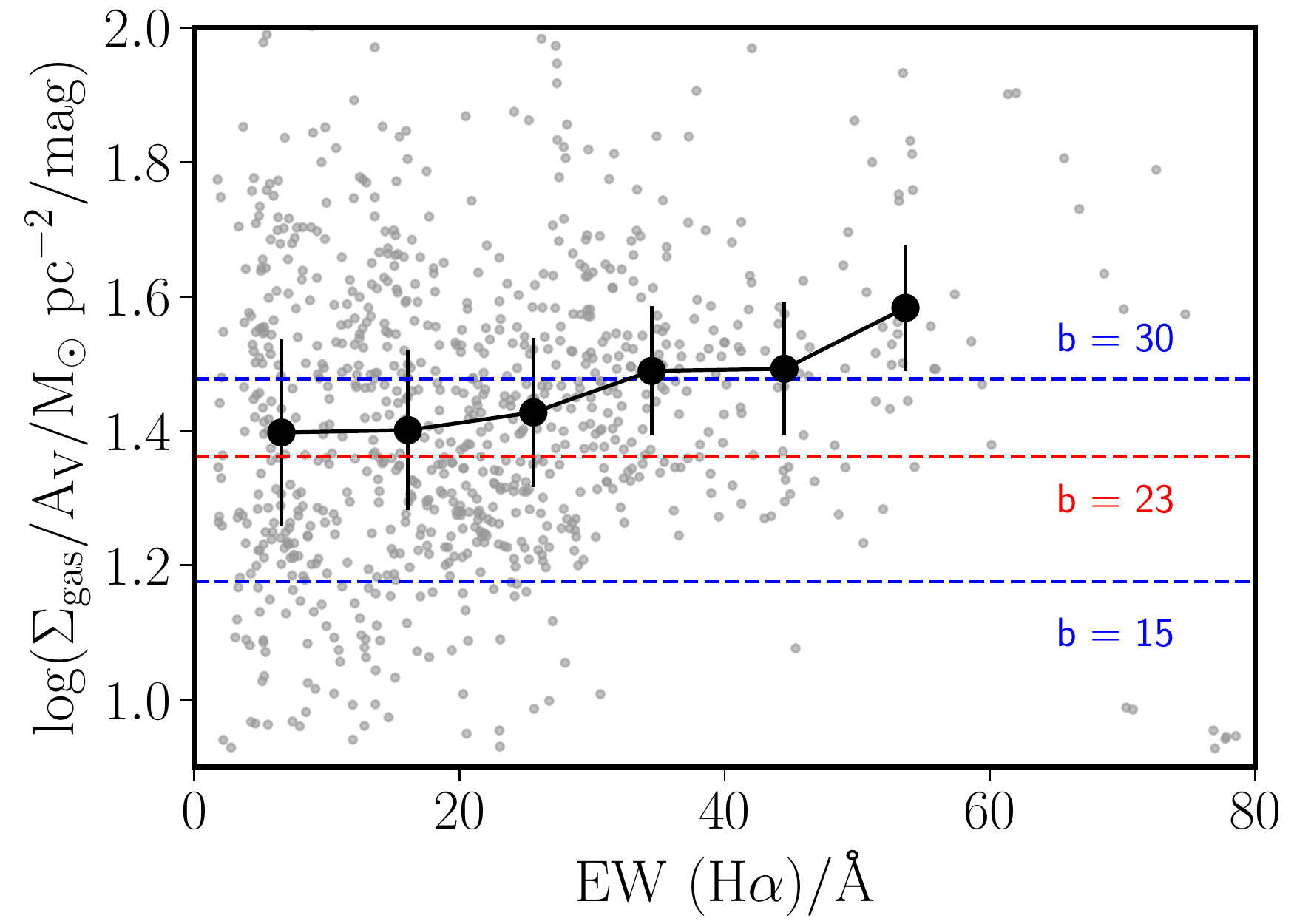} 
\includegraphics[width=0.98\linewidth]{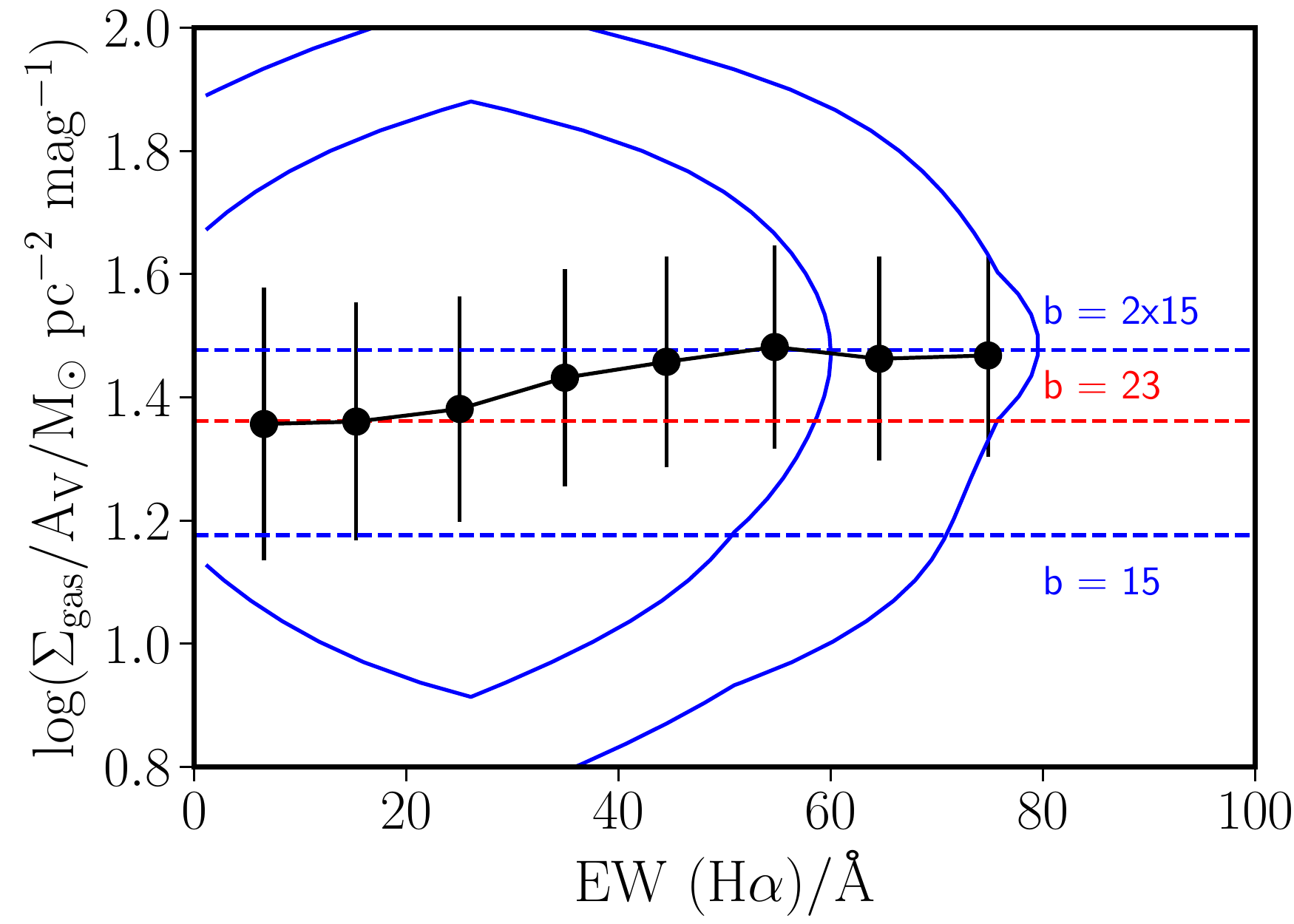} 
\caption{Comparison of the \Sgas-\av ratio with $\mathrm{EW(\ha)}$ for radial (top panel, gray points) and spaxel scales (bottom panel, blue contours). Black circles with errorbars represent the medians and standard deviations of the \Sgas-\av ratio in bins of 10 \AA\ of $\mathrm{EW(\ha)}$. As $\mathrm{EW(\ha)}$ increases, the  \Sgas-\av ratio increases reaching the value expected for an even-mixed geometry ($30\,\mathrm{M_{\odot}\,pc^{-2} \,mag^{-1}}$, upper blue dashed line).  }
\label{fig:EW}
\end{figure}

In Fig.~\ref{fig:EW} we plot this ratio against $\mathrm{EW(\ha)}$ for the radial (top panel) and spaxel (bottom panel) scales. For the radial study, we average the $\mathrm{EW(\ha)}$ in each annulus. For both scales, black points represent medians while error bars represent the standard deviation for the \Sgas-\av ratio in bins of width of $\mathrm{EW(\ha)}$ = 10\AA. In both panels, blue horizontal-dashed lines represent the values of the \mbox{\Sgas-\av} ratio expected for an obscuring screen and even-mix geometries \citep[$b~=~15$ and $30\,\mathrm{M_{\odot}\,pc^{-2} \,mag^{-1}}$,][]{Heiderman_2010} as well as the  ratio derived by \cite{Guever_2009} (red-dashed line with $b~=~23\,\,\mathrm{M_{\odot}\,pc^{-2}}$). We note that in average, regions with larger $\mathrm{EW(\ha)}$ tend to have larger \Sgas-\av ratios reaching a constant value of $\sim 30\,\mathrm{M_{\odot}\,pc^{-2} \,mag^{-1}}$. This is the ratio expected for a geometry in which dust is distributed both in the foreground and in the background of stars. These results suggest that for star-forming regions ($\mathrm{EW(\ha)} \gtrsim$  20 \AA) the \Sgas-\av relation is better represented by an even-mix geometry rather than for a single foreground dust-obscuring screen. On the other hand, for those regions with low $\mathrm{EW(\ha)}$ values although we are not probing a \Sgas-\av ratio close to a even-mix geometry we cannot rule out that this could be case. As we mention above, our method to determine \av relies in the detection of the \hb emission line which is quite difficult to measure for $\av > 3$. This may induce a bias towards those regions with low $\mathrm{EW(\ha)}$ having low \Sgas-\av ratios. It could also be the case that we are underestimating \av as our method is not sensitive to \av above 3 \citep[e.g.,][]{Liu_2013}. In any case, for those region that we are probing active star formation the prefer geometry seems to be a dust-mixed one.

\subsection{Validation of dust-inferred gas masses}
\label{sec:Comp}
\begin{figure}
\includegraphics[width=\linewidth]{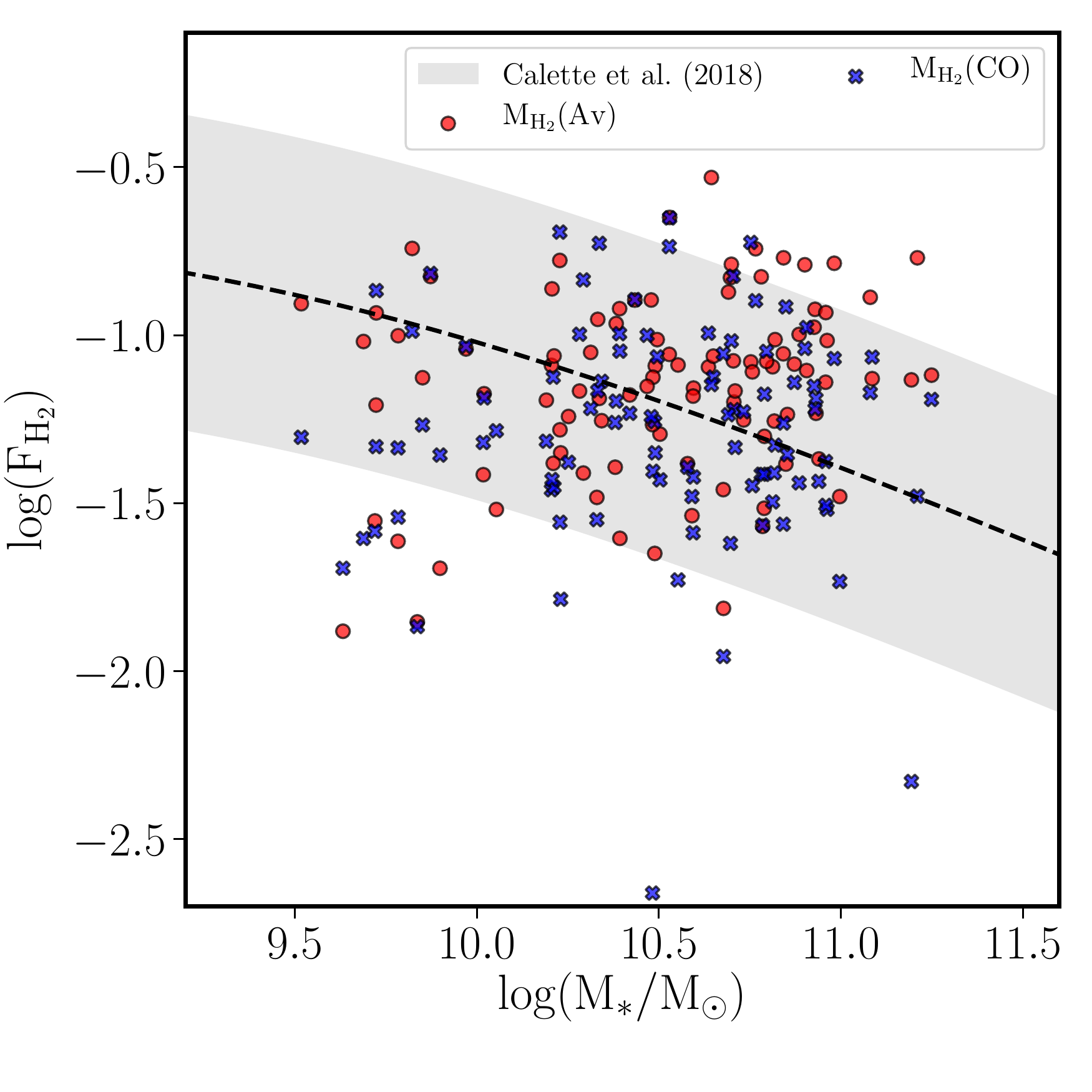}
\caption{Fraction of the total molecular gas mass (\mbox{${\rm F_{\htwo}} = {\rm M_{H_{2}}/ M_{\ast}}$}) vs the total stellar mass for our sample.  ${\rm F_{\htwo}}$ is derived using different indicators. The gray-shaded area and black dashed line represent the relation derived from the compilation presented by \protect\cite{Calette_2018}. Blue crosses represent the gas fraction derived from direct observations of CO (see details in Sec.~\ref{sec:CompInt}). ${\rm F_{\htwo}}$ derived from \av estimates (red circles) is similar to both the compiled relation and the values derived from direct CO observations.}
\label{fig:Int_comp}
\end{figure}
So far in this work we have studied the reliability of using dust attenuation derived from the  \ha / \hb ratio as a proxy for the amount of cold gas. The EDGE-CALIFA  data set allows us to perform such a study on a variety of spatial scales, from galaxy-integrated to spatially resolved scales. In this section, we provide a comparison between our estimates of the gas fraction using \av and compilations in the literature for both integrated and spatially resolved measurements. 

\subsubsection{Galaxy-integrated Measurements}
\label{sec:CompInt}
\label{sec:CompRes}
\begin{figure*}
    \includegraphics[width=0.48\linewidth]{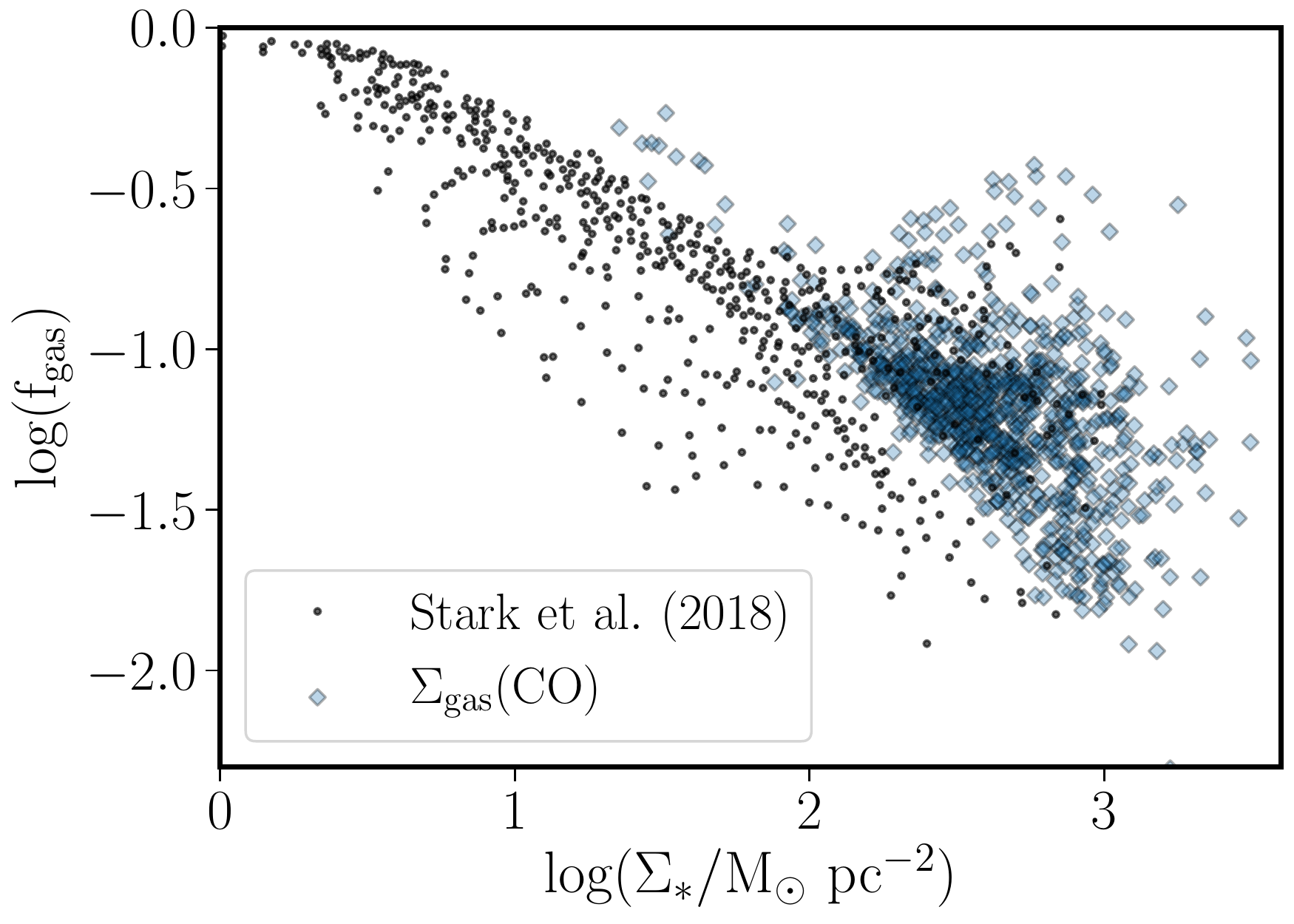} 
    \includegraphics[width=0.48\linewidth]{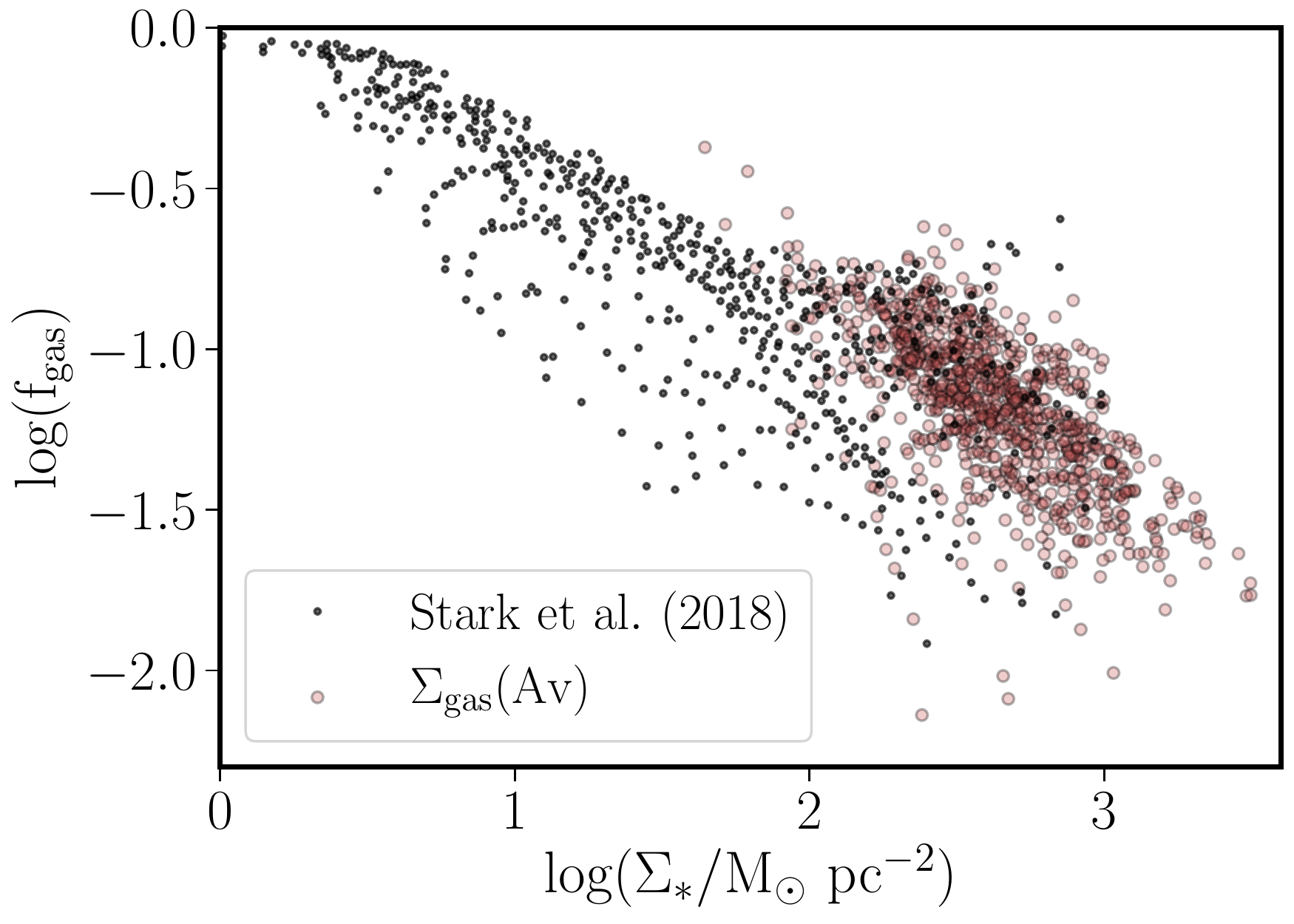}
\caption{Spatially resolved gas fraction (${\rm f_{gas}} = \Sgas/ \Sgas +\Sstar$ with  \Sgas = \Shtwo + \sighi ) as a function of \Sstar measured in radial bins. {\it Left panel:} Blue diamonds represent  ${\rm f_{gas}}$ with \Shtwo derived from the EDGE-CALIFA data set (see details in Sec.\ref{sec:Radial}).  {\it Right panel:} Red circles represent  ${\rm f_{gas}}$ with \Sgas derived using \av as a proxy for gas surface density as described in Sec.~\ref{sec:Radial}. The gray circles in both panels represent the  ${\rm f_{gas}}$ by Manga galaxies \protect\cite{Stark_2018}. The gas fraction relationship with \Sstar derived using \av follows (and extends) the relationship seen by \protect\cite{Stark_2018} at least as well as ${\rm f_{gas}}$  estimated from \Sgas.}
\label{fig:Res_comp}
\end{figure*}

In Fig.~\ref{fig:Int_comp} we present the scaling relation between total molecular gas fraction defined as ${\rm F_{\htwo}} = {\rm M_{H_{2}} / M_{\ast}}$ and stellar mass $M_{\ast}$. In this figure we used different estimates for the molecular mass. The red circles show the molecular gas fraction derived from the calibrator presented in Sec.~\ref{sec:Global} using as proxy the dust extinction. Blue crosses represent direct observations of CO for the EDGE-CALIFA galaxies \citep{Bolatto_2017}. We also compare these observations with a recent compilation in the literature of the trend of these two observables  \citep[see gray shaded area and the dashed-black line, ][]{Calette_2018}. The total stellar mass derived from this compilation assumed the initial mass function (IMF) from \cite{Chabrier_2003}, whereas the total stellar mass from CALIFA galaxies is derived using a \cite{Salpeter_1955} IMF.  We transform the relation derived from the compilation to agree with the stellar masses reported for the CALIFA sample. This figure shows that in general the molecular mass derived using \av\  is a good proxy for estimating the global molecular gas fraction. Red circles are in good agreement to the blue crosses. This is particularly useful when no direct observations of CO are available for a given set of galaxies. Most of the  ${\rm F_{\htwo}}$ derived using \av  are located within the stripe derived using a large compilation of different data sets. In comparison to other large cold gas surveys like the xCOLD GASS survey,  the values of ${\rm F_{\htwo}}$ in our sample derived using \av are similar to those derived using single-dish CO observations where the two samples overlap.

\subsubsection{Spatially Resolved Measurements}
We also perform a comparison of the gas content derived at radial scales with a sample of spatially resolved data reported in the literature. In Fig.~\ref{fig:Res_comp} we show the scaling relation between the gas fraction at local scales defined as ${\rm f_{gas}} = {\rm \Sigma_{gas} / [\Sigma_{gas} +\Sigma_{\ast}}$]  and the stellar mass density $\rm{\Sigma_{\ast}}$. In both panels the gray circles represent the data presented in \cite{Stark_2018}. They use a subsample of MaNGA galaxies \citep{Bundy_2015} with direct measurements of the cold gas component \citep{Leroy_2008}. The blue diamonds in the left panel represent the gas fraction derived using the EDGE-CALIFA CO radial measurements to trace $\rm{\Sigma_{H_{2}}}$ presented in Sec.~\ref{sec:Radial} and assuming that \sighi = 6 \msunperpcsq. First we note that our sample of galaxies probes larger values of $\rm{\Sigma_{\ast}}$ than those in  \cite{Stark_2018}. We also find that the values of ${\rm f_{gas}}$ derived from our sample agree very well with the trend observed by \cite{Stark_2018} at small values of $\rm{\Sigma_{\ast}}$; this is that the gas fraction decreases as the stellar mass density increases. In the right panel of Fig.~\ref{fig:Res_comp} we show in red circles the gas fraction derived using the $\rm{\Sigma_{gas}}$ obtained from the extinction using the calibrator presented in Sec.~\ref{sec:Radial}. As in the left-side plot, the gas fraction derived using extinction of \ha and \hb as a proxy is also in agreement with the trend observed in the direct measurements from \cite{Stark_2018}. In other words, this plot shows that the extinction is a reliable proxy for probing the gas fraction on local scales.

\subsection{Caveats}
\label{sec:IcoAvSen}
As we mention throughout this study, in order to determine a relation between \Sgas and \av it is necessary to consider the sensitivity of our observables.  Our ability to determine the optical extinction requires a reliable determination of the H$\beta$ flux. Therefore we cannot measure \av in regions where it is small enough that it is not affecting the \ha/\hb ratio, but even more importantly in those regions where extinction is large enough to suppress the H$\beta$ emission line. In contrast to Galactic studies \citep{Heiderman_2010, Lee_2018}, where they can probe regions with $\av > 10\,\,$mag, from the \ha/\hb ratio we can only probe regions with \av $\lesssim 3$ mag. On the other hand, the 3$\sigma$ sensitivity of our CO observations is \Shtwo $\sim$ 11~\msunperpcsq \citep{Bolatto_2017}.  This may imply a limitation in constraining the \Sgas-\av relation over a wider range of parameters. Nevertheless, we argue that for the parameters probed by the EDGE-CALIFA in both global and spatially-resolved scales (e.g., Figs \ref{fig:Int_comp}, \ref{fig:Res_comp}), the optical extinction derived from the \ha/\hb ratio is indeed a reliable proxy for estimating the amount of cold gas in extragalactic sources. In a future study, we will explore the \mbox{\Sgas-\av} relation using the optical extinction estimated from the fitting of the stellar continuum.

\section{Discussion}
\label{sec:Discussion}
\subsection{Previous CO and optical extinction measurements}
\label{sec:IcoAvRel}
In this work we present the observational relation between the gas mass surface density (\Sgas) and the optical extinction \av derived from the Balmer decrement for a sample of 103 galaxies included in the EDGE-CALIFA survey. The spatially resolved nature of this dataset allow us to determine this relation at integrated, radial and  kpc scales. In other words, we aim to provide a reliable proxy for direct observations of the gas surface density (via CO emission) using optical observables (via the optical extinction derived from the Balmer decrement). On theoretical grounds a relation between CO emission (\Ico) and optical extinction (\av) is expected. The amount of dust shielding between the CO-\mbox{$\rm C_{II}$} (or \hii-\hi) transition layer is almost constant \citep{Wolfire_2010, Lee_2015}. Similar relationships have been also been derived in photo-dissociation regions (PDR) calculations \citep[e.g,][]{Bell_2006} as well as in numerical simulations \citep[e.g.,][]{Glover_2011}

Observationally, the relation between gas and dust tracers has been studied on different scales and with different proxies. \cite{Lee_2015,Lee_2018} studied how the brightness of \Ico and the optical extinction derived from SED fitting of the infrared emission correlated in molecular clouds located in the Large Magellanic Cloud (LMC), Small Magallenic Cloud (SMC) and the Milky Way. Broadly speaking, they found that their empirical \Ico-\av relation agrees with the standard Galactic CO-to-\htwo\, conversion factor \citep[\xco = $2 \times 10^{20}$ \xcounits,][]{Bolatto_2008} with a mild dependence on the metallicity. Following \cite{Lee_2015} this conversion factor can be written as $\Ico/\av \sim 4.7\,~{\mbox{\rm K km s$^{-1}$ mag$^{-1}$}}$. From our analysis in Sec.~\ref{sec:Spaxels}, we find that this ratio for our sample corresponds to \Ico/\av = $5.4~{\mbox{\rm K km s$^{-1}$ mag$^{-1}$}}$. Although this value is larger by 15\% than the \Ico/\av  ratio expected from the CO-to-\htwo\, conversion factor, we argue that we are using a more heterogeneous set of CO emitting regions and moreover we are averaging these properties over areas larger than those in studies in the local Volume. It would important to probe this ratio on sub-kpc scales in extragalactic objects. This potentially could be achieved by comparing the optical extinction from MUSE datacube observations with \Ico data from ALMA millimeter observations. Despite these caveats, it is clear that in the absence of direct millimeter observations it is still possible to use optical proxies such as the \av derived from the Balmer decrement to gauge the content of the cold gas in the ISM.  

Recently, \cite{Concas_2019} present a similar analysis as the one presented in this study for global properties by determining the relation between the Balmer decrement and the total molecular mass (${\rm M_{H_{2}}(CO)}$) for a sample of 222 star-forming galaxies included in the xCOLD GASS survey. They indicate that scatter is reduced when highly-inclined galaxies are excluded. We compare our integrated measurements with the best relation derived from their data set. To do this, we transform our global measurements of \av to Balmer decrement. Our data is in agreement with their best estimates of the ${\rm M_{H_{2}}(CO)}$ - Balmer decrement relation, with most of the data points lying between their best fit for all the star-forming galaxies and the line derived using only low-inclination targets. These results reinforce the idea that the \av derived from the Balmer decrement is a reliable proxy for estimating the content of gas in large surveys which have no direct observations in the millimeter.

\subsection{The physical interpretation of the scatter of the gas-extinction relation}
\label{sec:IcoAvRel}

In Sec.~\ref{sec:Results} we note that the scatter of the \mbox{\Sgas-\av } relation increases as we observe this relation at smaller physical scales (from global, radial and spaxel scales; see Figs.~\ref{fig:Int_Av_SCO}, \ref{fig:Rad_Av_SCO}, and \ref{fig:spaxel_Av_SCO}, respectively). We suggest that this could be a consequence of probing smaller scales.  The smaller the physical scale, the more complex is the geometrical relation between the amount of gas/dust and stars \citep[e.g.,][]{Liu_2013,Tacchella_2018}. In other words, when we measure these observables at larger physical scales we are averaging out the complexity in the physical distribution of the stars, dust and gas.  

As mentioned above, the simplest dust/star geometry -- foreground dust screen -- yields the smallest \Sgas/\av ratio \citep[15 \sigavUnits, ][]{Heiderman_2010} whereas a mixed geometry would yield twice the ratio accounting for dust and gas in background and in the foreground of the stars \citep[e.g.,][]{Nordon_2013}. Although the best fit of the \mbox{\Sgas-\av} relation for the probed physical scales for the EDGE-CALIFA galaxies is closer to the ratio reported by \cite{Guever_2009} (i.e., \Sgas/\av = 23 \sigavUnits) because their scatter, our reported relations cover a wide range of ratios. 

In Sec.~\ref{sec:EW}, we note that this ratio correlates with the $\mathrm{EW(\ha)}$ -- a tracer of star-formation activity -- at radial and spaxel scales. In average, the larger the $\mathrm{EW(\ha)}$ the larger the \Sgas/\av ratio. In particular, we note that for $\mathrm{EW(\ha)} > 20$ \AA\ (i.e., actively star-forming regions) this ratio tend to plateau towards $\sim$ 30 \sigavUnits. This indicates that for star-forming galaxies/regions the most likely distribution of the dust/gas and stars corresponds to a dust-star mixed geometry \citep[e.g.,][]{Wuyts_2011, Genzel_2013}. Finally, we indicate that the smallest spatial scale that we are able to probe in this sample of nearby galaxies is of the order of kpc which is very large compared with studies of molecular clouds within the Galaxy. It is important to probe whether the \Sgas-\av relation holds in the sub-kpc regime in extragalactic objects representative of the local Universe.  

\subsection{The impact of metallicity on the gas-extinction relation.}
\label{sec:ImpactOH}
\begin{figure}
\includegraphics[width=\linewidth]{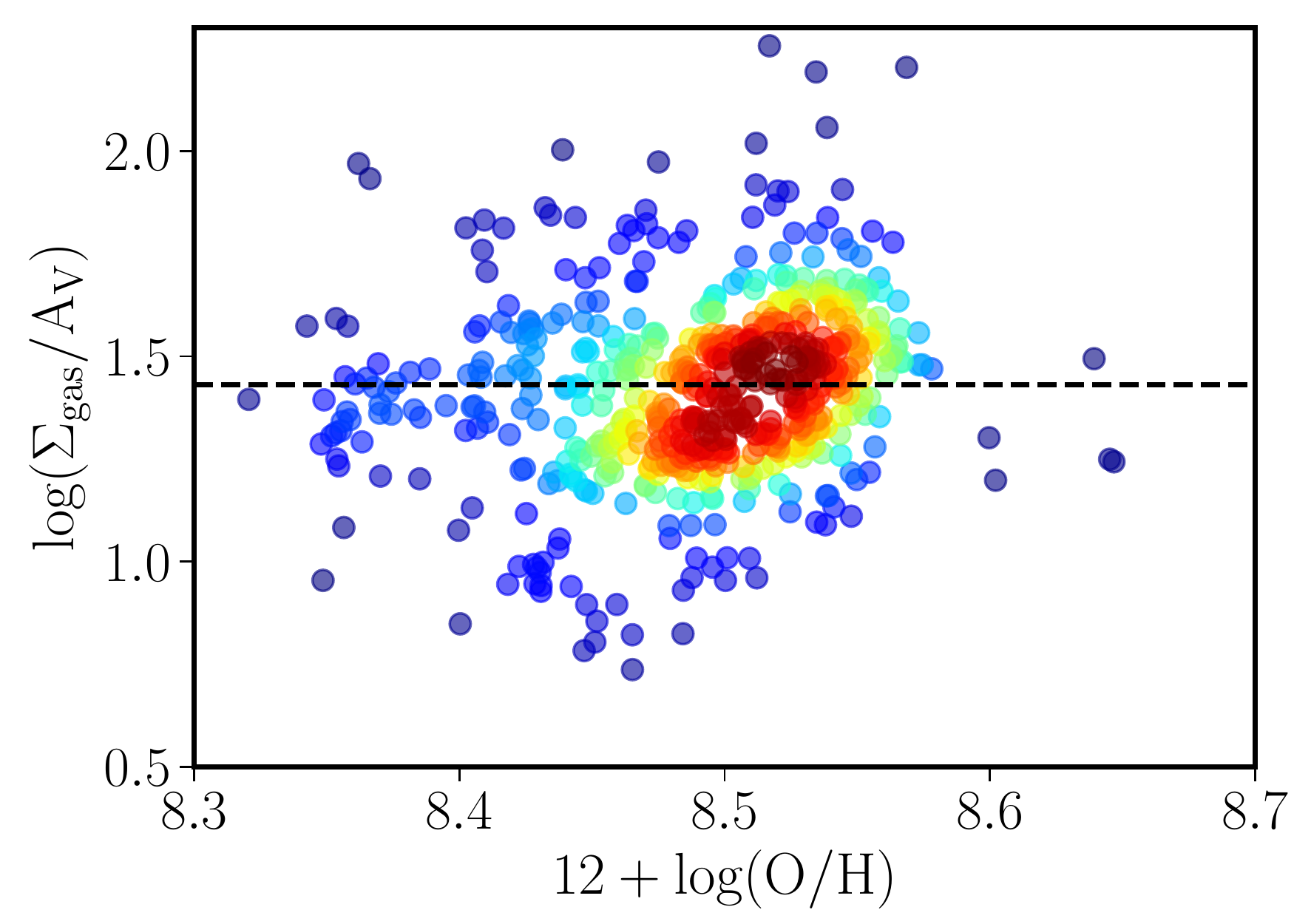}
\caption{The ratio \Sgas/\av in units of $\log(\msunperpcsqpermag)$ plotted vs the ionized gas metallicity for selected star-forming radial bins. The dashed line represents the best fit from the radial \Sgas-\av relation (i.e., 27 \msunperpcsqpermag, see Sec~\ref{sec:Radial} for details). The impact of metallicity on the relationship between \Sgas and \av is small.} 
\label{fig:Ras_OH}
\end{figure}
The \xco\, factor is expected to depend on the gas metallicity \citep[e.g.,][]{Bolatto_2008, Boquien_2013}. To test how the metallicity could potentially impact the determination of the gas mass from the extinction, we plot in Fig.~\ref{fig:Ras_OH} the logarithm of the ratio between \Sgas and \av  versus the gas metallicity ($\rm{12+\log(O/H)}$) only for those radial bins derived in Sec.~\ref{sec:Radial} considered as star-forming. To determine which bins are star-forming, we select those  with emission line ratios \nii/\ha and $\rm{[O{\small III}]}$/\hb  classified as star-forming. This is regions with ratios below the \cite{Kauffmann_2003_BPT} demarcation line in the classical BPT diagnostic diagram \citep{BPT_1981}. As well as those regions with  $\mathrm{EW(\ha)} > 6$\AA. We determine the metallicity using the O3N2 calibrator derived by \cite{Marino_2013}. Thus, for this analysis we use 593 bins from the 893 bins described in Sec~\ref{sec:Radial}. The dynamical range in metallicity probed is rather narrow, covering only $\sim$0.3 dex. We find a rather flat distribution of this ratio with respect to the gas metallicity. To highlight the flatness of this relation, we plot a dashed-line in Fig.~\ref{fig:Ras_OH} representing the best fit of the \Sgas-\av relation derived in Sec.~\ref{sec:Radial}. The  values center on this dashed-line, reinforcing the idea that, at least for these galaxies,  metallicity does not seem to play an important role in shaping the \Sgas-\av relation. We suspect that the small impact of  metallicity relation could also be induced by the fact that we are not probing low-mass galaxies in our sample. The EDGE-CALIFA survey covers a range of relatively massive galaxies \cite[i.e., $10^{10} - 10^{12}$ \msun; see details in][]{Bolatto_2017}; therefore we are not able to probe the gas content, optical extinction and metallicity for low-mass galaxies. We consider that further studies including a deeper analysis  between spatially resolved millimetre and optical data such as the EDGE-CALIFA survey are necessary to probe the low-mass regime and quantify the impact of metallicity on the determination of the gas surface density on local scales. 

\subsection{Scaling relations using the Gas-Extinction calibrator}
\label{sec:CompRes}
\begin{figure*}
\includegraphics[width=0.46\textwidth]{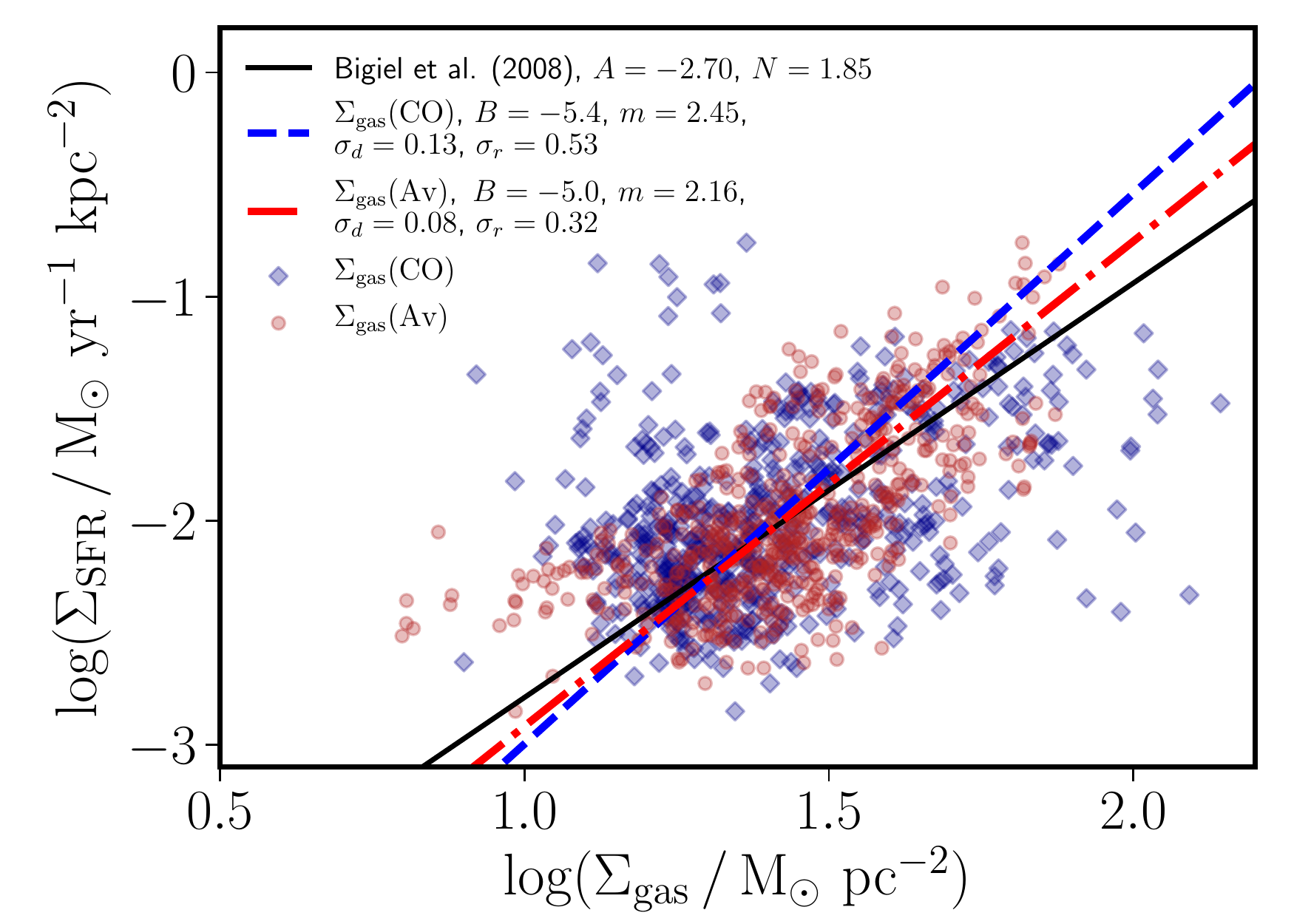} 
\includegraphics[width=0.46\textwidth]{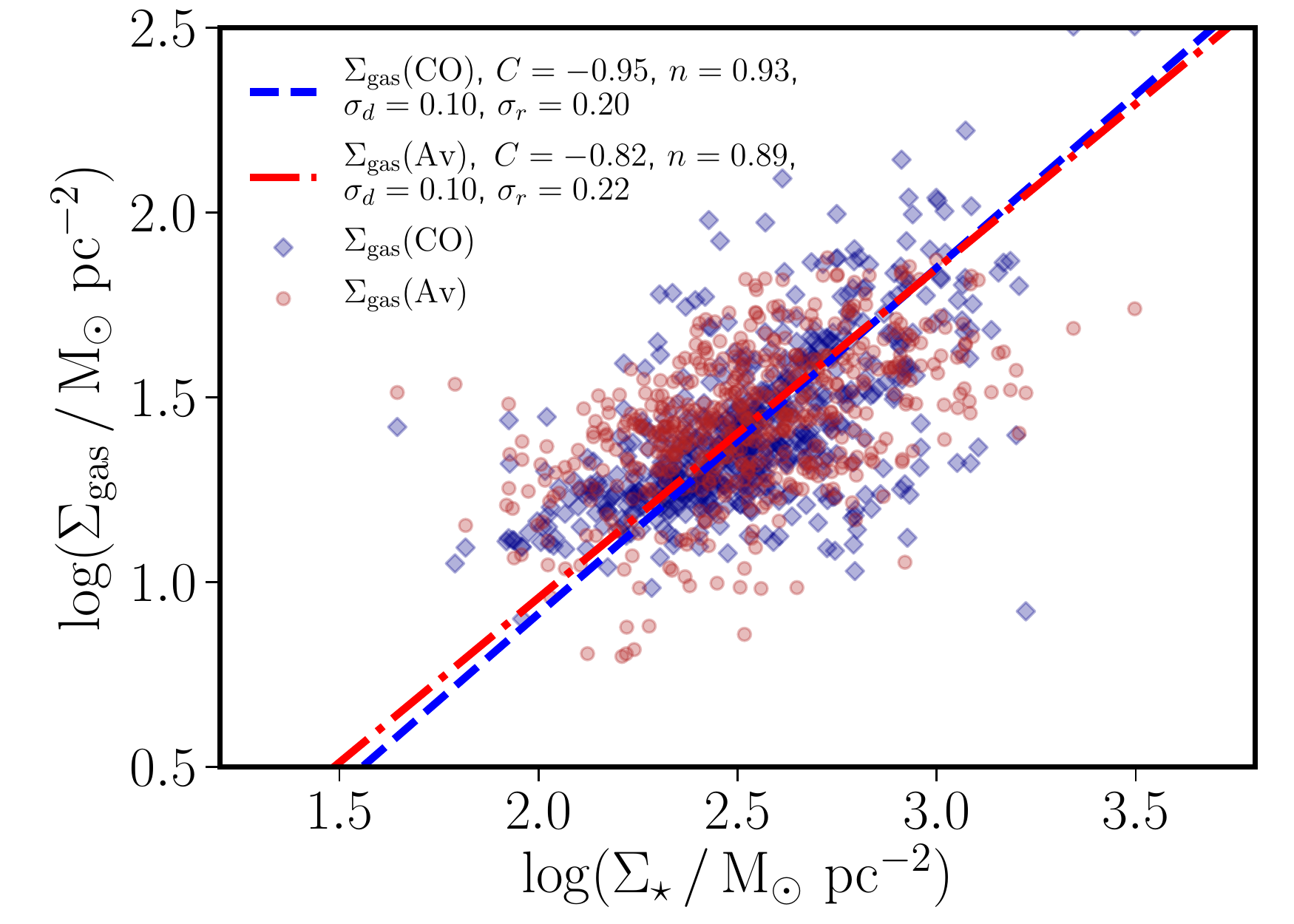} 
\caption{Scaling relation for \Sgas using different estimates at radial scales. In both panels blue diamonds represent \Sgas determined using CO to estimate \Shtwo whereas red circles represent \Sgas determined using \av (see details in Sec.~\ref{sec:Radial}). Left panel shows the star formation rate density (\Ssfr) vs \Sgas. The black solid line represents the star formation law derived from observations by \protect\cite{Bigiel_2008}.  The right panel shows the \Sstar-\Sgas scaling relation. In both panels blue dashed and red dot-dashed lines represent the best ODR fitting for these scaling relation using CO and \av to estimate \Sgas, respectively (see details in Sec.~\ref{sec:CompRes}). For each of these fits we also include the dispersion from orthogonal distance ($\sigma_d$) and the dispersion from the residuals ($\sigma_r$). The similarity of the fitting and the residuals using either calibrator indicate that \av is a reliable calibrator to estimate \Sgas in  the absence of direct observations of the cold gas component.}
\label{fig:Ras_KS}
\end{figure*}
In this section we discuss whether our proxy  of \Sgas derived from  optical extinction is sufficiently reliable to obtain spatially resolved  scaling relations. In particular, we focus on the star-formation law  and the \Sstar-\Sgas relations on radial scales. The star-formation law  \citep[also known as the Kennicutt-Schmidt Law,][]{Kennicutt_1998} describes how the star formation rate density (\Ssfr) tightly correlates with \Sgas. We determine \Ssfr in each radial bin considered as star-forming following the same selection procedure as described in Sec.\ref{sec:ImpactOH}. The star formation rate (SFR) in each bin is derived using the relation between SFR and \ha luminosity presented in Eq.~\ref{eq:SFR_Ha}. Then, \Ssfr in each radial bin is obtained by dividing the SFR by the area of the radial bin. 

In the left panel of Fig.~\ref{fig:Ras_KS} we present the star-formation law using two different estimates of \Sgas. Blue diamonds represent  \Sgas derived using the CO radial measurements as a proxy for \Shtwo as presented in Sec.~\ref{sec:Radial} and assuming that \sighi is 6 \msunperpcsq. Red circles represent \Sgas using the best fit of the relation \Sgas-\av presented in Sec.~\ref{sec:Radial}. The black line shows the average of the best fit of the Kennicutt-Schmidt law on logarithm scales for seven spiral galaxies presented by \cite{Bigiel_2008} (cf., Eq.~2). We note that even though we are using the same index ($N=1.85$) as \cite{Bigiel_2008}, we are using the lowest coefficient reported by their uncertainties in ($A = -2.64$). 

We note that both estimates of the star formation law using different proxies for \Sgas are in excellent agreement with each other. Even more, when we compare our estimates with the estimate presented by \cite{Bigiel_2008}, we find good agreement in both the slope of the law as well as the coefficient. We further provide the best fit of this relation using and ODR fitting. We fit the equation \mbox{$\log(\Ssfr) = B + m\, \log(\Sgas)$} to the data. In the legend of the left panel of Fig.~\ref{fig:Ras_KS} we show the values from the fit for $B$ and $m$ parameters from both \Sgas estimates. The slopes of the fitted relations are slightly steeper than those derived from \citep{Bigiel_2008} with $m =$  2.54 and 2.16, for \Sgas (CO) and \Sgas (\av), respectively. On the other hand, the intercept with the \mbox{$\log(\Ssfr)$} axis is smaller than the one derived by \citep{Bigiel_2008} with $B =$ -5.4 and \mbox{-4.9}, respectively. Both slopes derived from our ODR fitting are in agreement with the one derived by \citep{Bigiel_2008} within the reported uncertainties. 

We also compute the scatter of the relations by measuring the standard deviation of the residuals of the data points with respect to the best fitted lines using two quantities: measuring ({\it i}) the orthogonal distance of the data points with respect to the best fitted line ($\sigma_d$) and ({\it ii}) the difference between the observed and expected \Ssfr ($\sigma_r$). For both of these parameters we find that \Sgas derived from \av has a smaller scatter than \Sgas from CO measurements ($\sigma_d$ = 0.08 dex; $\sigma_r$ = 0.32 dex vs. $\sigma_d$ = 0.13 dex; $\sigma_r$ = 0.53 dex). As a validation test, we also derive $\sigma_r$ with respect to the \cite{Bigiel_2008} best relation using the \Sgas derived from CO. We find a similar scatter as the one derived from the ODR best fitting but somewhat smaller with $\sigma_r$ = 0.43 dex. Our results may suggest that using \av to estimate \Sgas may lead to smaller scatter in the estimates of star formation law. However, we caution that this may not be the case since we are calibrating \Sgas from the best relation between \av and \Sgas derived using CO, this means that we are not including the scatter of the \Sgas-\av relation to the star-forming law presented here which evidently increases the scatter in this relation. We investigate those galaxies hosting annuli with outlying high \Ssfr/\Sgas. We find that these annuli are located within highly inclined galaxies and merging galaxies. To conclude, this analysis suggests that \av is a reliable proxy to trace the gas content in the context of the star formation law. Although it is beyond the scope of this paper to provide a detailed study of the star formation law, \cite{Bolatto_2017} note that the star formation law derived only for the molecular component for the EDGE-CALIFA sample is in excellent agreement with those derived in other CO surveys \citep[e.g., HERACLES survey][]{Leroy_2013}. 

Another important scaling relation is \Sgas vs \Sstar. In the right panel of  Fig.~\ref{fig:Ras_KS} we plot this relation using our CO and \av to estimate \Sgas. The data points for both estimates of \Sgas are in very good agreement with each other. As above, we perform an ODR fitting to both datasets using the equation \mbox{$\log(\Sstar) = C + n\, \log(\Sgas)$}. Our analysis shows that this relation can be described either using \av or CO as \Sgas estimator. Both estimates of \Sgas reproduce very similar results in the parameters of the best fitting (C = -0.92 and -0.88; n = 0.90 and 0.89 for \Sgas estimated using CO and \av, respectively). Even more, the observed scatter is very similar for both estimates of \Sgas ($\sigma_d$ = 0.10; $\sigma_r$ = 0.20,  $\sigma_d$ = 0.09; $\sigma_r$ = 0.22, for \Sgas estimated using CO and \av, respectively). Similar scaling relations have been derived using ALMA data in a sample of 14 MaNGA star-forming galaxies \citep{Lin_2019}. To conclude, we consider that our estimates of \Sgas from optical extinction to be a reliable proxy for use in order to measure the star formation law in star-forming galaxies. 

\section{Summary and Conclusions}
\label{sec:Conclusions}

Thanks to the EDGE-CALIFA survey we are able to determine an empirical relation between the gas surface density \Sgas (\Sgas = \Shtwo+\sighi) and optical extinction \av inferred from the \ha/\hb ratio in a sample of 103 galaxies in the nearby Universe. The angular-resolved nature of the EDGE (in the millimeter) and CALIFA (in the optical) surveys allows us to determine this relation on global, radial, and kpc (spaxel) scales. We summarize the main  results of this study as follows:
\begin{enumerate}
\itemsep0em
\item We measure the best relation between \Sgas and \av at global, radial, and spaxel scales (see Figs.~\ref{fig:Int_Av_SCO}, \ref{fig:Rad_Av_SCO}, and~\ref{fig:spaxel_Av_SCO}).  We find similar  relations across these different spatial scales, obtaining $\Sgas~(\msunperpcsq)~\sim~26~\times~ \av ({\rm mag})$. This is in agreement with previous studies \citep[e.g.,][]{Guever_2009}.

\item We find that even though the best fit relation for the \Sgas-\av relation is similar on the different spatial scales, the scatter of the relation increases as smaller scales are probed. We argue that this is an indication of the complexity in geometries or physical distribution of the stars, gas and dust on kpc or sub-kpc scales \citep[e.g.,][]{Liu_2013}. We note that on average for actively star forming annuli/spaxels (i.e, with $\mathrm{EW(\ha)} \gtrsim$  20 \AA, see Fig.\ref{fig:EW}) the \Sgas/\av ratio is closer to the expected value for a star-dust mixed geometry ($\sim$ 30 \sigavUnits). 

\item  We do not find a significant trend between the \Sgas/\av ratio and the ionized gas metallicity on radial scales for galaxy properties probed by the EDGE-CALIFA survey (we note however that our range of metallicities is rather small ~ 0.2 dex). It would be worth exploring this relation in galaxies of lower stellar mass.   

\item We derive the gas fractions using the best value for \Sgas derived from \av and compare these fractions with compilations in the literature for the global and radial scales (see Figs.~\ref{fig:Int_comp} and \ref{fig:Res_comp}). We also derive the star-forming and \Sgas-\Sstar\ scaling relations for radial scales (see Fig.~\ref{fig:Ras_KS}) using \Sgas from CO direct measurements and  estimates from \av. We find excellent agreement between the two \Sgas estimators.  These comparisons show that \av is a reliable proxy to use in the absence of direct estimates of \Sgas. 
\end{enumerate}

In conclusion, using the galaxies sampled by the EDGE-CALIFA survey we determine that the optical extinction derived from the Balmer decrement is a reliable tracer of the gas component measured from the CO observations. This observational calibrator is quite useful in particular for optical studies of large samples of galaxies using IFS observations which lack spatially resolved observations of the cold gas component \citep[e.g.,][]{BB_2018,Sanchez_2018}.

\section*{Acknowledgements}
\addcontentsline{toc}{section}{Acknowledgements}

J.K.B-B would like to thank to the referee for the constructive comments. J.K.B-B, S.F.S. thank CONACYT grant CB285080 and funding from the PAPIIT-DGAPA-IA101217 (UNAM) project. This study makes use of data from the EDGE (\url{www.astro.umd.edu/EDGE}) and CALIFA (\url{http://califa.caha.es}) surveys and numerical values from the HyperLeda database (\url{http://leda.univ-lyon1.fr}). Support for CARMA construction was derived from the Gordon and Betty Moore Foundation, the Kenneth T. and Eileen L. Norris Foundation, the James S. McDonnell Foundation, the Associates of the California Institute of Technology, the University of Chicago, the states of California, Illinois, and Maryland, and the NSF. CARMA development and operations were supported by the NSF under a cooperative agreement and by the CARMA partner universities. This research is based on observations collected at the Centro Astron\'{o}mico Hispano-Alem\'{a}n (CAHA) at Calar Alto, operated jointly by the Max-Planck Institut f\"{u}r Astronomie (MPA) and the Instituto de Astrofisica de Andalucia (CSIC).



\bibliographystyle{mnras}
\bibliography{main}


\bsp	
\label{lastpage}
\end{document}